\documentclass[12pt]{article}
\usepackage{amsmath,amssymb}
\newcommand{\eqdef}{\stackrel{\rm def}{=}}
\newcommand{\n}{\nonumber \\}
\newcommand{\bm}{\boldsymbol}
\renewcommand{\theequation}{\arabic{section}.\arabic{equation}}

\allowdisplaybreaks[3]

\begin{document}

\baselineskip=20pt

\newfont{\elevenmib}{cmmib10 scaled\magstep1}
\newcommand{\preprint}{
    \begin{flushleft}
      \elevenmib Yukawa\, Institute\, Kyoto\\
    \end{flushleft}\vspace{-1.3cm}
    \begin{flushright}\normalsize  \sf
      DPSU-06-1\\
      YITP-06-23\\
     {\tt quant-ph/0605215}\\
       May 2006
    \end{flushright}}
\newcommand{\Title}[1]{{\baselineskip=26pt
    \begin{center} \Large \bf #1 \\ \ \\ \end{center}}}
\newcommand{\Author}{\begin{center}
    \large \bf Satoru~Odake${}^a$ and Ryu~Sasaki${}^b$ \end{center}}
\newcommand{\Address}{\begin{center}
      $^a$ Department of Physics, Shinshu University,\\
      Matsumoto 390-8621, Japan\\
      ${}^b$ Yukawa Institute for Theoretical Physics,\\
      Kyoto University, Kyoto 606-8502, Japan
    \end{center}}
\newcommand{\Accepted}[1]{\begin{center}
    {\large \sf #1}\\ \vspace{1mm}{\small \sf
Accepted for Publication}
    \end{center}}

\preprint
\thispagestyle{empty}
\bigskip\bigskip\bigskip

\Title{Unified Theory of Annihilation-Creation Operators for 
Solvable (`Discrete') Quantum Mechanics}
\Author

\Address

\begin{abstract}
The annihilation-creation operators $a^{(\pm)}$ are defined as 
the positive/negative frequency parts of the exact Heisenberg operator
solution for the `sinusoidal coordinate'. Thus $a^{(\pm)}$ are
hermitian conjugate to each other and the relative weights of various
terms in them are solely determined by the energy spectrum.
This unified method applies to most of the solvable quantum mechanics of
single degree of freedom including those belonging to the `discrete'
quantum mechanics.
\end{abstract}

\section{Introduction}
\label{intro}
\setcounter{equation}{0}

The annihilation and creation operators are probably the most
basic and important tools in quantum mechanics.
Modern quantum physics is almost unthinkable without them.
About eighty years after its birth, the list of exactly solvable
systems in quantum mechanics \cite{susyqm} is quite long now, including 
those of the so-called `discrete' quantum mechanics \cite{os4,os3}.
One natural question is that if these exactly solvable quantum 
mechanical systems also possess the algebraic solution method embodied
in the annihilation-creation operators.
We will answer the question in the affirmative and give a 
{\em unified dynamical\/} derivation of the annihilation-creation operators
for most of the solvable quantum mechanics of single degree of freedom
including those belonging to the `discrete' quantum mechanics.

The method is quite simple and elementary. One identifies a special
function $\eta(x)$ of the space-coordinate $x$, which undergoes
`sinusoidal motion' at the classical and quantum levels
\eqref{clasol}, \eqref{quantsol}. 
The latter is simply the exact Heisenberg operator solution
for $\eta(x)$.
The function $\eta$ is the argument of the orthogonal polynomial 
\eqref{special} constituting the eigenfunctions of the system. 
The positive/negative frequency parts of the exact Heisenberg 
operator solution of the `sinusoidal coordinate'
$\eta(x)$ \eqref{acdefs0}, \eqref{acdefs} give the annihilation-creation
operators. This is essentially the same recipe as used by Heisenberg
for solving the harmonic oscillator in matrix mechanics.
To the best of our knowledge, the `sinusoidal coordinate' was first
introduced in a rather broad sense for general 
(not necessarily solvable) potentials as a useful means for coherent 
state research by Nieto and Simmons \cite{nieto}. 

By similarity transformation in terms of the ground state wavefunction,
the results of the present paper will be translated to those of the 
corresponding orthogonal polynomials.
In particular, the exact Heisenberg operator solution of the 
`sinusoidal coordinate' 
corresponds to the so-called {\em structure relation\/} \cite{koorn}
for the orthogonal polynomials. 
Our method provides the unified derivation of the structure relations
for the Askey-Wilson, Wilson, continuous dual Hahn, continuous Hahn
and Meixner-Pollaczek polynomials \cite{koeswart} based on the
Hamiltonian principle.
These polynomials are the eigenfunctions of the various `discrete'
quantum mechanical systems \cite{os4,os3} and they are the deformations
of the Jacobi, Laguerre and Hermite polynomials \cite{szego}.

This paper is organised as follows. The general theory of 
the annihilation-creation operators is explained with typical examples;
one from the ordinary quantum mechanics and two from the `discrete'
version in section two.
Further explicit results are presented in section three.
They include the (symmetric) P\"oshl-Teller, symmetric Rosen-Morse,
Morse and $x^2+1/x^2$ potentials on top of the five examples belonging
to the `discrete' quantum mechanics mentioned above. Section four is
for a summary and comments. In Appendix {\bf A}, the necessary and
sufficient condition for the existence of the `sinusoidal coordinate'
is analysed within the context of ordinary quantum mechanics.
It turns out that those potentials  having the `sinusoidal coordinate'
are all {\em shape invariant\/} \cite{genden}.
In Appendix {\bf B}, interpretation of the annihilation-creation
operators within the framework of shape invariance is given. 
It is the mechanism underlying the solvability of all
systems considered in this paper.
Appendix {\bf C} gives various definitions of the 
orthogonal polynomials, hypergeometric functions and their $q$-analogues
\cite{szego,koeswart}.

\section{General theory with typical examples}
\label{gentheory}
\setcounter{equation}{0}

The purpose of the present paper is to present a 
{\em  unified dynamical\/} theory of 
{\bf annihilation-creation} operators.  
It is applicable to most of the {\em exactly solvable\/} quantum 
mechanical systems of one degree of freedom, 
including the so-called `discrete' quantum mechanics which are
certain deformation of the solvable quantum mechanics \cite{os4,os3}.
They satisfy certain {\em difference equations\/} instead of the
second order differential equations. Generalisation to the
systems of many degrees of freedom will be discussed elsewhere.
The restriction to the solvable quantum systems is rather trivial 
and inevitable, since a system is obviously solvable 
if it possesses explicitly defined annihilation and creation 
operators and any one single eigenstate to work on. 
Then the entire set of exact eigenstates are easily and 
concretely generated.

Except for the simple harmonic oscillator,  
which gives probably the only
so far universally accepted example of the annihilation-creation
operators, there are quite a wide variety of proposed annihilation 
and creation operators in the literature \cite{coherents}. Historically
most of these  annihilation-creation operators are connected to the
so-called  {\em algebraic theory of coherent states\/}, which are
usually defined as eigenstates of {\em annihilation operators\/}. 
Therefore, for a given  potential or a quantum Hamiltonian, 
there could be as many coherent states as the definitions of the 
annihilation operators.

Our new unified definition of the annihilation-creation operators 
is, on the contrary, based on the dynamical properties of a special 
coordinate, the `sinusoidal coordinate' shared by a class of 
solvable dynamical systems discussed in this paper.
A quantum mechanical system with a self-adjoint Hamiltonian
$\mathcal{H}$ is solvable (or solved) if the entire set of its
energy eigenvalues $\{\mathcal{E}_n\}$ and the corresponding 
eigenvectors $\{\phi_n\}$, $n=0,1,\ldots$ are known:
\begin{equation}
  \mathcal{H}\phi_n=\mathcal{E}_n\phi_n,\quad n=0,1,\ldots,\, .
\end{equation}
As is well-known a quantum Hamiltonian (together with its `discrete' 
analogue) has in general discrete as well as continuous spectrum.
In this paper we will concentrate on the discrete energy levels only,
either finite or infinite in number.
Then, because of the one-dimensionality, the eigenvalues are not
degenerate
\begin{equation}
  \mathcal{E}_0 <\mathcal{E}_1 < \cdots,
\end{equation}
and the eigenvectors have finite norms $||\phi_n||^2=N_n^2<\infty$.
(When normalised vectors are needed we denote them by adding a hat,
$\{\hat{\phi}_n\eqdef \phi_n/N_n\}$, $||\hat{\phi}_n||=1$.)
This is the solution in the  Schr\"odinger picture and the
eigenvectors $\{\phi_n\}$ are usually expressed as functions
$\{\phi_n(x)\}$ of the space-coordinate $x$. 
For the majority of the solvable quantum systems, the $n$-th 
eigenfunction has the following general structure \cite{susyqm}:
\begin{equation}
  \phi_n(x)=\phi_0(x) P_n(\eta(x))
  \label{special}
\end{equation}
in which $\phi_0(x)$ is the {\em ground state\/} wavefunction.
It has no nodes and we may choose it to be always real and positive.
The second factor $P_n(\eta(x))$ is a polynomial of degree $n$ in 
a real variable $\eta$. We also take  $P_n(\eta)$ as real
and use a convention $P_{-1}(\eta)=0$.
Reflecting the orthogonality theorem of the eigenvectors of a 
self-adjoint Hamiltonian, $\{P_n(\eta)\}$ form 
{\em orthogonal polynomials\/} with respect to a weight function 
(measure)
\begin{equation}
  \phi_0(x)^2dx\propto w(\eta)d\eta.
\label{measure}
\end{equation}
Throughout this paper we follow the definition and notation of
Szeg\"o's book \cite{szego} for the classical orthogonal polynomials 
and the review by Koekoek and Swarttouw \cite{koeswart} for the 
Askey-scheme of hypergeometric orthogonal polynomials and its
$q$-analogue,  which are deformations of the classical orthogonal
polynomials.

There are certain exactly solvable quantum systems, for example,
the one-dimensional Kepler problem, etc, for which the general
form of eigenfunction \eqref{special} does not hold. For them,
the present unified theory does not apply. See Appendix {\bf A}.

Our main claim is that this $\eta(x)$ undergoes a 
`{\em sinusoidal motion\/}' under the given Hamiltonian $\mathcal{H}$,
at the classical as well as quantum level, by mimicking the simple 
harmonic oscillator. This fact is the basis of our dynamical and 
unified  definition of the annihilation-creation operators. 
To be more specific, at the classical level we have
\begin{equation}
  \{\mathcal{H},\{\mathcal{H},\eta\}\}=-\eta\,
  R_0(\mathcal{H})-R_{-1}(\mathcal{H})
  \label{twopoi}
\end{equation}
in which the canonical Poisson bracket relations are defined 
for the canonical coordinate $x$, its conjugate momentum $p$ and 
for any functions $A(x,p)$ and $B(x,p)$ as:
\begin{equation}
  \{x,p\}=1,\quad \{x,x\}=\{p,p\}=0,\quad
  \{A,B\}=\frac{\partial A}{\partial x}\frac{\partial B}{\partial p}
  -\frac{\partial A}{\partial p}\frac{\partial B}{\partial x}.
\end{equation}
The two coefficients $R_0$ and $R_{-1}$ are, in general, 
polynomials in the Hamiltonian $\mathcal{H}$.
The effect of $R_{-1}$ is to shift the origin of $\eta(x)$ by a 
quantity (possibly) depending on $\mathcal{H}$.
It is convenient to introduce a shifted sinusoidal coordinate
$\tilde{\eta}(x)$
\begin{equation}
  \tilde{\eta}(x)\eqdef\eta(x)+R_{-1}(\mathcal{H})/R_{0}(\mathcal{H})
  \qquad\Bigl(\Rightarrow
  \{\mathcal{H},\{\mathcal{H},\tilde{\eta}\}\}=-\tilde{\eta}\,
  R_0(\mathcal{H})\Bigr).
\end{equation}
The relation \eqref{twopoi} would allow to evaluate the multiple 
Poisson brackets of $\eta$ with $\mathcal{H}$ easily:
\begin{equation}
  {\rm ad}\,\mathcal{H}\,\eta\eqdef\{\mathcal{H},\eta\},\quad 
  ({\rm ad}\,\mathcal{H})^2\eta=\{\mathcal{H},\{\mathcal{H},\eta\}\},\quad
  ({\rm ad}\,\mathcal{H})^n\eta
  =\{\mathcal{H},({\rm ad}\,\mathcal{H})^{n-1}\eta\},
\end{equation}
which leads to a simple sinusoidal time-evolution:
\begin{align}
  \tilde{\eta}(x;t)&=\sum_{n=0}^\infty\frac{(-t)^n}{n!}
  ({\rm ad}\,\mathcal{H})^n\tilde{\eta},\n
  &=-\{\mathcal{H},\tilde{\eta}\}_0\,
  \frac{\sin\bigl[t\sqrt{R_0(\mathcal{H}_0)}\,\bigr]}
  {\sqrt{R_0(\mathcal{H}_0)}}
  +\tilde{\eta}(x)_0\cos\bigl[t\sqrt{R_0(\mathcal{H}_0)}\,\bigr]
  \label{clasol0}.
\end{align}
In the original variable it reads
\begin{align}
  \eta(x;t)&=-\{\mathcal{H},\eta\}_0\,
  \frac{\sin\bigl[t\sqrt{R_0(\mathcal{H}_0)}\,\bigr]}
  {\sqrt{R_0(\mathcal{H}_0)}}
  -R_{-1}(\mathcal{H}_0)/R_{0}(\mathcal{H}_0)\n
  &\quad
  +\bigl(\eta(x)_0+R_{-1}(\mathcal{H}_0)/R_{0}(\mathcal{H}_0)\bigr)
  \cos\bigl[t\sqrt{R_0(\mathcal{H}_0)}\,\bigr],
  \label{clasol}
\end{align}
in which $\eta(x)_0$ ( $\tilde{\eta}(x)_0$) and 
$\{\mathcal{H},\eta\}_0$ are the initial values (at $t=0$) of these 
variables and $\mathcal{H}_0$ denotes the value of the Hamiltonian 
(the energy) for these initial data.
In general, the frequency of the simple oscillation
$\sqrt{R_0(\mathcal{H}_0)}$ can depend on the initial data. 
This is the reason why we call $\eta(x)$ the `sinusoidal coordinate' 
avoiding the more appealing but misleading ``harmonic coordinate".

At the quantum level with the canonical commutation relations
\begin{equation}
  [x,p]=i,\quad [x,x]=[p,p]=0,\qquad \hbar\equiv1,
\end{equation}
the formula corresponding to \eqref{twopoi} reads
\begin{equation}
  [\mathcal{H},[\mathcal{H},\eta]\,]=\eta\,R_0(\mathcal{H})
  +[\mathcal{H},\eta]\,R_1(\mathcal{H})+R_{-1}(\mathcal{H}).
  \label{twocom}
\end{equation}
In other words, the multiple commutators of $\mathcal{H}$ with
$\eta$ form a closed algebra at level two.
Here, the quantum coefficients $R_0$ and $R_{-1}$ could differ
from the classical ones by quantum corrections.  
But we use the same symbols since there is no risk of confusion.
Obviously $R_1(\mathcal{H})$ is the quantum effect.
As in the classical case, the multiple commutators of $\eta$ with
$\mathcal{H}$
\begin{equation}
  {\rm ad}\,\mathcal{H}\,\eta\eqdef[\mathcal{H},\eta],\quad 
  ({\rm ad}\,\mathcal{H})^2\eta=[\mathcal{H},[\mathcal{H},\eta]\,],\quad
  ({\rm ad}\,\mathcal{H})^n\eta
  =[\mathcal{H},({\rm ad}\,\mathcal{H})^{n-1}\eta],
\end{equation}
can be easily evaluated from \eqref{twocom}.
This leads to the {\em exact operator solution\/} in the Heisenberg
picture:
\begin{align}
   e^{it\mathcal{H}}\eta(x)e^{-it\mathcal{H}}
  &=\sum_{n=0}^\infty\frac{(it)^n}{n!}({\rm ad}\,\mathcal{H})^n\eta,\n
  &=[\mathcal{H},\eta(x)]
  \frac{e^{i\alpha_+(\mathcal{H})t}-e^{i\alpha_-(\mathcal{H})t}}
  {\alpha_+(\mathcal{H})-\alpha_-(\mathcal{H})}
  -R_{-1}(\mathcal{H})/R_{0}(\mathcal{H})\n
  &\quad
  +\bigl(\eta(x)+R_{-1}(\mathcal{H})/R_0(\mathcal{H})\bigr)
  \frac{-\alpha_-(\mathcal{H})e^{i\alpha_+(\mathcal{H})t}
  +\alpha_+(\mathcal{H})e^{i\alpha_-(\mathcal{H})t}}
  {\alpha_+(\mathcal{H})-\alpha_-(\mathcal{H})},
  \label{quantsol}
\end{align}
in which the two ``frequencies" $\alpha_\pm(\mathcal{H})$ are
\begin{gather}
  \alpha_\pm(\mathcal{H})=\bigl(R_1(\mathcal{H})\pm
  \sqrt{R_1(\mathcal{H})^2+4R_0(\mathcal{H})}\,\bigr)/2, \\
  \alpha_+(\mathcal{H})+\alpha_-(\mathcal{H})=R_1(\mathcal{H}),
  \quad
  \alpha_+(\mathcal{H})\alpha_-(\mathcal{H})=-R_0(\mathcal{H}).
  \label{freqpm}
\end{gather}
If the quantum effects are neglected, {\em i.e.} $R_1\equiv0$ and
$\mathcal{H}\to\mathcal{H}_0$ (in the r.h.s.),
we have $\alpha_+=-\alpha_-=\sqrt{R_0(\mathcal{H}_0)}$, the above
Heisenberg operator solution reduces to the classical one \eqref{clasol}
in terms of the quantum-classical correspondence:
\begin{equation}
  [A,B]/i\hbar\to \{A,B\},\quad (\hbar\to0).
\end{equation}
The above exact operator solution looks slightly simpler if the shifted
sinusoidal coordinate is used
\begin{equation}
  e^{it\mathcal{H}}\tilde{\eta}(x)e^{-it\mathcal{H}}
  =[\mathcal{H},\tilde{\eta}(x)]
  \frac{e^{i\alpha_+(\mathcal{H})t}-e^{i\alpha_-(\mathcal{H})t}}
  {\alpha_+(\mathcal{H})-\alpha_-(\mathcal{H})}
  +\tilde{\eta}(x)
  \frac{-\alpha_-(\mathcal{H})e^{i\alpha_+(\mathcal{H})t}
  +\alpha_+(\mathcal{H})e^{i\alpha_-(\mathcal{H})t}}
  {\alpha_+(\mathcal{H})-\alpha_-(\mathcal{H})}.
  \label{quantsol0}
\end{equation}

Like the exact classical solution \eqref{clasol}, the exact quantum 
solution \eqref{quantsol} contains all the dynamical information of 
the quantum system. 
One can, for example, determine the entire discrete spectrum
$\{\mathcal{E}_n\}$ by following Heisenberg and Pauli's arguments for the
harmonic oscillator and the Hydrogen atom. Let us first note that the
ground state energy $\mathcal{E}_0=0$ is known explicitly, because of our
choice of the factorised form  of the exactly solvable Hamiltonian
(see examples below):
\begin{equation}
  \mathcal{H}=\mathcal{A}^\dagger \mathcal{A}/2,\quad
  \mathcal{A}\phi_0=0 \Rightarrow
  \mathcal{H}\phi_0=0,\quad
  \mathcal{E}_0=0.
  \label{factham}
\end{equation}
Let us apply \eqref{quantsol} to the $n$-th eigenvector $\phi_n$:
\begin{align}
  &\quad e^{it(\mathcal{H}-\mathcal{E}_n)}\eta(x)\,\phi_n\n
  &=\Bigl([\mathcal{H},\eta(x)]\phi_n+\bigl(-\eta(x)\alpha_-(\mathcal{E}_n)+
  R_{-1}(\mathcal{E}_n)/\alpha_+(\mathcal{E}_n)\bigr)\phi_n\Bigr)
  e^{i\alpha_+(\mathcal{E}_n)t}/
  \bigl(\alpha_+(\mathcal{E}_n)-\alpha_-(\mathcal{E}_n)\bigr)\n
  &\quad
  +\Bigl(-[\mathcal{H},\eta(x)]\phi_n+\bigl(\eta(x)\alpha_+(\mathcal{E}_n)-
  R_{-1}(\mathcal{E}_n)/\alpha_-(\mathcal{E}_n)\bigr)\phi_n\Bigr)
  e^{i\alpha_-(\mathcal{E}_n)t}/
  \bigl(\alpha_+(\mathcal{E}_n)-\alpha_-(\mathcal{E}_n)\bigr)\n
  &\quad
  -\bigl(R_{-1}(\mathcal{E}_n)/R_0(\mathcal{E}_n)\bigr)\phi_n.
  \label{nthreeterm}
\end{align}
Since the r.h.s. has only two different time dependence except for
the constant term, the l.h.s. can only have two non-vanishing matrix 
elements when sandwiched by $\phi_m$, except for the obvious $\phi_n$ 
corresponding to the constant term. In accordance with the general
structure of the eigenfunctions \eqref{special}, they are $\phi_{n\pm1}$:
\begin{equation}
  \langle\phi_m|\eta(x)|\phi_n\rangle=0,\quad \mbox{for}
  \quad m\neq n\pm1,n.
  \label{simple3}
\end{equation}
This imposes the following conditions on the energy eigenvalues
\begin{equation}
  \mathcal{E}_{n+1}-\mathcal{E}_{n}=\alpha_+(\mathcal{E}_n),\quad
  \mathcal{E}_{n-1}-\mathcal{E}_{n}=\alpha_-(\mathcal{E}_n).
  \label{cond1}
\end{equation}
Likewise we obtain the `hermitian conjugate' conditions
\begin{equation}
  \mathcal{E}_{n}-\mathcal{E}_{n-1}=\alpha_+(\mathcal{E}_{n-1}),\quad
  \mathcal{E}_{n}-\mathcal{E}_{n+1}=\alpha_-(\mathcal{E}_{n+1})
  \label{cond2}
\end{equation}
relating these three neighbouring eigenvalues. These overdetermined
conditions \eqref{cond1}, \eqref{cond2} and $\mathcal{E}_0=0$ determine
the entire energy spectrum $\{\mathcal{E}_n\}$ completely for each
Hamiltonian. The consistency of the procedure requires that the second
term on r.h.s. of \eqref{nthreeterm} should vanish when applied to the
ground state $\phi_0$:
\begin{equation}
  -[\mathcal{H},\eta(x)]\phi_0+\bigl(\eta(x)\alpha_+(0)
  -R_{-1}(0)/\alpha_-(0)\bigr)\phi_0=0,
\end{equation}
which could be interpreted as the equation determining the ground 
state eigenvector $\phi_0$ in the Heisenberg picture.
If the number of the discrete levels is finite ($M+1$) a corresponding 
condition must be met that the first term on r.h.s. of \eqref{nthreeterm} 
should not belong to the Hilbert space of normalisable vectors
when applied to the highest discrete level eigenvector $\phi_M$:
\begin{equation}
  \bigl|\bigr|
  [\mathcal{H},\eta(x)]\phi_M+\bigl(-\eta(x)\alpha_-(\mathcal{E}_M)
  +R_{-1}(\mathcal{E}_M)/\alpha_+(\mathcal{E}_M)\bigr)\phi_M
  \bigl|\bigr|=\infty.
  \label{norm}
\end{equation}

As is clear by now, \eqref{quantsol} and \eqref{simple3} are the 
physical embodiment of the {\bf three term recursion relations} 
satisfied by any orthogonal polynomial of single variable:
\begin{equation}
  \eta P_n(\eta)=A_n P_{n+1}(\eta)+B_n P_n(\eta)+C_n P_{n-1}(\eta).
  \label{P3term}
\end{equation}
The coefficients $A_n$, $B_n$ and $C_n$ are
also real for a real polynomial $P_n(\eta)$. 
For the systems treated in this paper, that is those having the 
general structure of the eigenvectors \eqref{special}, this implies 
the three term recurrence relations of the eigenfunctions
\begin{equation}
  \eta(x)\phi_n(x)=A_n \phi_{n+1}(x)+B_n \phi_n(x)+C_n \phi_{n-1}(x).
  \label{phi3term}
\end{equation}
At the same time the above arguments and the treatment of the harmonic
oscillator by Heisenberg clearly show that the operator coefficient of
$e^{it\alpha_-(\mathcal{H})}$ on the r.h.s. of the Heisenberg operator
solution \eqref{quantsol} is the annihilation operator, that is, acting
on $\phi_n$ it produces a state $\phi_{n-1}$.
Likewise, the operator coefficient of
$e^{it\alpha_+(\mathcal{H})}$ of the Heisenberg operator
solution \eqref{quantsol} is the creation operator. 
Thus we arrive at a {\em dynamical and unified  definition\/} of 
the {\bf annihilation-creation operators}:
\begin{align}
  &e^{it\mathcal{H}}\eta(x)e^{-it\mathcal{H}}
  =a^{(+)}(\mathcal{H},\eta)e^{i\alpha_+(\mathcal{H})t}
  +a^{(-)}(\mathcal{H},\eta)e^{i\alpha_-(\mathcal{H})t}
  -R_{-1}(\mathcal{H})/R_{0}(\mathcal{H}),
  \label{acdefs0}\\
  &a^{(\pm)}=a^{(\pm)}(\mathcal{H},\eta)\n
  &\phantom{a^{(\pm)}}\eqdef\Bigl(\pm[\mathcal{H},\eta(x)]
  \mp\bigl(\eta(x)+R_{-1}(\mathcal{H})/R_0(\mathcal{H})\bigr)
  \alpha_\mp(\mathcal{H})\Bigr)\!\bigm/\!
  \bigl(\alpha_+(\mathcal{H})-\alpha_-(\mathcal{H})\bigr).
  \label{acdefs}
\end{align}
When acting on the eigenvector $\phi_n$, they read
\begin{equation}
  a^{(\pm)}\phi_n(x)=\frac{\pm1}{\mathcal{E}_{n+1}-\mathcal{E}_{n-1}}
  \Bigl([\mathcal{H},\eta(x)]+(\mathcal{E}_n-\mathcal{E}_{n\mp 1})\eta(x)
  +\frac{R_{-1}(\mathcal{E}_n)}{\mathcal{E}_{n\pm 1}-\mathcal{E}_n}\Bigr)
  \phi_n(x).
  \label{gen1}
\end{equation}

\bigskip
Before going to the detailed discussion of the annihilation-creation operators
for various Hamiltonians in section \ref{typexam} and section \ref{3rd},
let us analyse annihilation-creation operators in a more general context.
A minimal requirement for annihilation-creation operators is the
following:
\begin{align}
  \mbox{(0)} : &\mbox{
  annihilation-creation operators map $\phi_n$ to $\phi_{n-1}$ and 
  $\phi_{n+1}$ (up to an overall}\qquad\qquad\n
 &\mbox{ constant), respectively.}\nonumber
\end{align}
It should be stressed that there is no a priori principle for fixing the
normalisation of the operators. 
Sometimes it is convenient to introduce the annihilation-creation operators
with a different normalisation
\begin{equation}
  a^{\prime(\pm)}\eqdef a^{(\pm)}
  \bigl(\alpha_+(\mathcal{H})-\alpha_-(\mathcal{H})\bigr)
  =\pm[\mathcal{H},\eta(x)]
  \mp\bigl(\eta(x)+R_{-1}(\mathcal{H})/R_0(\mathcal{H})\bigr)
  \alpha_\mp(\mathcal{H}),
  \label{ac'defs}
\end{equation}
which gives
\begin{equation}
  a^{\prime(\pm)}\phi_n(x)=
  \pm\Bigl([\mathcal{H},\eta(x)]+(\mathcal{E}_n-\mathcal{E}_{n\mp 1})\eta(x)
  +\frac{R_{-1}(\mathcal{E}_n)}{\mathcal{E}_{n\pm 1}-\mathcal{E}_n}\Bigr)
  \phi_n(x).
  \label{gen'1}
\end{equation}
In the r.h.s. the coefficients of the operator $\eta(x)$ and the identity
operator depend on $n$ in general.

The annihilation-creation operators of the {\bf harmonic oscillator}
have several remarkable properties:
\label{acprop}
\begin{align}
  \mbox{(\romannumeral1)} : & \mbox{
  Annihilation/creation operator is the positive/negative frequency 
  part of}\n
  & \mbox{the Heisenberg operator of the `sinusoidal coordinate'.}\n
  \mbox{(\romannumeral2)} : & \mbox{
  (annihilation operator)$^{\dagger}$=(creation operator).}\n
  \mbox{(\romannumeral3)} : & \mbox{
  $\mathcal{H}={\rm const.}\times$
  (creation operator)(annihilation operator).}\nonumber
\end{align}
The first property (\romannumeral1) is the principle leading to our
unified  definition of the 
annihilation-creation operators as in \eqref{acdefs0}--\eqref{acdefs}. 
Next we show that they are hermitian conjugate to each other.
That is, they satisfy the property (\romannumeral2),  too. 
By using the three-term recursion relation,  we obtain
\begin{equation}
  e^{it\mathcal{H}}\eta e^{-it\mathcal{H}}\phi_n
  =e^{it(\mathcal{E}_{n+1}-\mathcal{E}_n)}A_n\phi_{n+1}
  +B_n\phi_n+e^{it(\mathcal{E}_{n-1}-\mathcal{E}_n)}C_n\phi_{n-1}.
\end{equation}
Comparing this with  \eqref{acdefs0} and \eqref{acdefs},
we arrive at
\begin{equation}
a^{(+)}\phi_n=A_n\phi_{n+1},\quad a^{(-)}\phi_n=C_n\phi_{n-1},\quad
  R_{-1}(\mathcal{E}_n)/R_0(\mathcal{E}_n)=-B_n.
  \label{acphi=}
\end{equation}
Here use is made of the facts
\begin{align}
  \alpha_+(\mathcal{E}_n)&=\mathcal{E}_{n+1}-\mathcal{E}_n,\quad
  \alpha_-(\mathcal{E}_n)=\mathcal{E}_{n-1}-\mathcal{E}_n,\nonumber\\
  R_1(\mathcal{E}_n)&=\mathcal{E}_{n+1}+\mathcal{E}_{n-1}-2\mathcal{E}_n,
  \quad
  R_0(\mathcal{E}_n)=-(\mathcal{E}_{n+1}-\mathcal{E}_n)
  (\mathcal{E}_{n-1}-\mathcal{E}_n),
\end{align}
and that $\alpha_{\pm}(\mathcal{H})$ and $R_i(\mathcal{H})$ ($i=1,0,-1$)
are hermitian.
Hermitian conjugate of $a^{(-)}$ is
\begin{equation}
  a^{(-)\dagger}
  =\bigl(\alpha_+(\mathcal{H})-\alpha_-(\mathcal{H})\bigr)^{-1}
  \Bigl([\mathcal{H},\eta(x)]+\alpha_+(\mathcal{H})
  \bigl(\eta(x)+R_{-1}(\mathcal{H})/R_0(\mathcal{H})\bigr)
  \Bigr),
\end{equation}
and its action on $\phi_n$ is
\begin{align}
  a^{(-)\dagger}\phi_n&=
  \bigl(\alpha_+(\mathcal{H})-\alpha_-(\mathcal{H})\bigr)^{-1}
  \Bigl(\mathcal{H}\eta\phi_n-\eta\mathcal{E}_n\phi_n
  +\alpha_+(\mathcal{H})
  \bigl(\eta\phi_n+R_{-1}(\mathcal{E}_n)/R_0(\mathcal{E}_n)\phi_n\bigr)
  \Bigr)\n
  &=\bigl(\alpha_+(\mathcal{H})-\alpha_-(\mathcal{H})\bigr)^{-1}
  \Bigl((\mathcal{E}_{n+1}-\mathcal{E}_n)A_n\phi_{n+1}
  +(\mathcal{E}_{n-1}-\mathcal{E}_n)C_n\phi_{n-1}\n
  &\qquad\qquad\qquad\qquad\qquad\quad
  +\alpha_+(\mathcal{H})(A_n\phi_{n+1}+C_n\phi_{n-1})
  \Bigr)\n
  &=\bigl(\alpha_+(\mathcal{H})-\alpha_-(\mathcal{H})\bigr)^{-1}
  (\mathcal{E}_{n+2}-\mathcal{E}_n)A_n\phi_{n+1}\n
  &=A_n\phi_{n+1}=a^{(+)}\phi_n.
\end{align}
Therefore $a^{(\pm)}$ are hermitian conjugate to each other,
$a^{(-)\dagger}=a^{(+)}$. 
This also means that
\begin{equation}
e^{-it\alpha_{-}(\mathcal{H})}a^{(+)}=a^{(+)}e^{it\alpha_{+}(\mathcal{H})},
\quad
e^{-it\alpha_{+}(\mathcal{H})}a^{(-)}=a^{(-)}e^{it\alpha_{-}(\mathcal{H})},
\end{equation}
reflecting the obvious hermiticity of the l.h.s. of \eqref{acdefs0}.
Note that $a^{\prime(\pm)}$ are not hermitian conjugate to each other,
$a^{\prime(-)\dagger}\neq a^{\prime(+)}$, in general.

In the special case of the equi-spaced spectrum $\mathcal{E}_n=an$
($a$ : constant), to which many interesting examples belong 
including the harmonic oscillator and its deformation, we have
$\alpha_{\pm}(\mathcal{H})=\pm a$, $R_1(\mathcal{H})=0$,
$R_0(\mathcal{H})=a^2$ and
\begin{eqnarray}
  2a a^{(\pm)}=a^{\prime(\pm)}=\pm [\mathcal{H},\eta(x)]+
  a\eta(x)+R_{-1}(\mathcal{H})/a.
\end{eqnarray}
For the simplest harmonic oscillator $\mathcal{H}=(p+ix)(p-ix)/2$, 
we have $\eta(x)=x$, $R_0=1$, $R_1=R_{-1}=0$ and $[\mathcal{H},x]=-ip$ 
and $2a^{(+)}=x-ip$, $2a^{(-)}=x+ip$, which differ from the 
conventional ones by a factor $\sqrt{2}$.

\bigskip
In contrast to the above two properties, the third the property 
(\romannumeral3) of the annihilation-creation operators is achieved by
a very specific modification of the definition as follows:
\begin{equation}
  a^{\prime\prime(-)}\eqdef a^{(-)}f(\mathcal{H}),\quad
  a^{\prime\prime(+)}\eqdef a^{\prime\prime(-)\dagger}
  =f(\mathcal{H})a^{(+)},
\end{equation}
where $f(\mathcal{H})=f(\mathcal{H})^{\dagger}$ is an as yet 
unspecified function of $\mathcal{H}$.
Then we have
\begin{equation}
  a^{\prime\prime(+)}a^{\prime\prime(-)}\phi_n=
  f(\mathcal{E}_n)^2A_{n-1}C_n\phi_n.
\end{equation}
If $f(\mathcal{H})$ is chosen to satisfy
\begin{equation}
  f(\mathcal{E}_n)^2=\mathcal{E}_n/(A_{n-1}C_n)\ \bigl(\ \eqdef g(n)\bigr),
\end{equation}
then we obtain 
\begin{equation}
  \mathcal{H}=a^{\prime\prime(+)}a^{\prime\prime(-)}.
\end{equation}
Such an operator $f(\mathcal{H})$ function can be constructed as 
$f(\mathcal{H})=\sqrt{g(\mathcal{N})}$, where the number (or level) 
operator $\mathcal{N}$ is defined by
\begin{equation}
  \mathcal{N}\phi_n=n\phi_n.
  \label{opN}
\end{equation}
This operator $\mathcal{N}$ can be expressed in terms of the
Hamiltonian, for example,
\begin{alignat}{2}
  \mathcal{E}_n&=an\ \ &\Rightarrow\ \ \mathcal{N}&=\mathcal{H}/a,\\
  \mathcal{E}_n&=an^2+bn\ \ &\Rightarrow\ \ \mathcal{N}
  &=\bigl(\sqrt{4a\mathcal{H}+b^2}-b\bigr)/(2a),\\
  \mathcal{E}_n&=a(q^{-n}-1)(1-bq^n)\ \ &\Rightarrow\ q^{\mathcal{N}}
  &=\bigl(\mathcal{H}/a+b+1-\sqrt{(\mathcal{H}/a+b+1)^2-4b}\,\bigr)
  /(2b),
\end{alignat}
where $a,b,q$ are constant ($b>0$ in the second equation, $b<1$ in 
the third equation).
This $a^{\prime\prime(\pm)}$ satisfies the property (\romannumeral2)
by definition, but the property (\romannumeral1) becomes ugly or 
unnatural in general.

\bigskip
Let us close the general theory by a brief discussion of the coherent states.
One definition of the coherent state $\psi$ is  the eigenvector of the
annihilation operator (AOCS, Annihilation Operator Coherent State):
\begin{equation}
  a^{(-)}\psi=\lambda\psi,\quad \lambda\in\mathbb{C}.
  \label{cohedef}
\end{equation}
In terms of the simple parametrisation
$\psi=\sum_{n=0}^\infty c_n \phi_n(x)$ (with $c_0=1$ as 
normalization) and the formula \eqref{acphi=}, we arrive at
\begin{equation}
  \psi=\psi(\lambda,x)=\phi_0(x)\sum_{n=0}^{\infty}
  \frac{\lambda^n}{\prod_{k=1}^nC_n}\cdot P_n(\eta(x)).
  \label{cohe}
\end{equation}
For the equi-spaced spectrum $\mathcal{E}_n=an$ ($a$ : constant), 
the coherent state has the property of temporal stability
\begin{equation}
  e^{it\mathcal{H}}\psi(\lambda,x)=\psi(e^{iat}\lambda,x).
\end{equation}
It should be remarked that the concrete form of the AOCS depends on the 
specific normalisation of the annihilation operator.
For the annihilation operators $a^{\prime(-)}$, $a^{\prime\prime(-)}$ and
others, we denote the corresponding coherent states as $\psi'$,
$\psi''$, etc.
Which coherent state is useful depends on the physics of the system.

\subsection{Some typical examples}
\label{typexam}

Now let us look at typical examples \cite{susyqm,os4,os3} to show the
actual content of our new unified theory of annihilation-creation
operators. In their pioneering work, Nieto and Simmons \cite{nieto}
treated four solvable cases, those discussed in sections \ref{sutpot},
\ref{calo}, \ref{soliton} and \ref{morse}. Some of our results were
reported in \cite{nieto}.
Here and throughout this paper we put the dimensionfull
quantities  as unity, including the Planck's constant.

\subsubsection{$1/\sin^2x$ potential, or symmetric P\"oschl-Teller potential}
\label{sutpot}

The first example has the  $1/\sin^2x$  potential,
which is the one-body case of the well-known Sutherland model
\cite{calsut,kps}. This provides  the simplest example of the
annihilation-creation  operators depending on $n$. The corresponding
coherent state
\eqref{sutcoherent1} or \eqref{sutcoherent2} had not yet been known, 
to the best of our knowledge.
The system is confined in a finite interval, say $(0,\pi)$ and 
it has an infinite number of discrete eigenstates. 
Although this potential is a special ($g=h$) case of the P\"oschl-Teller
potential discussed in section \ref{poshtell}, it merits separate
analysis. The Hamiltonian, the eigenvalues and the eigenfunctions
are as follows:
\begin{align}
  \mathcal{H}&\eqdef(p-ig\cot x)(p+ig\cot x)/2,\quad
  \bigl(\,\stackrel{{\rm Q.M.}}{\Longrightarrow}\,
  2\mathcal{H}+g^2=p^2+g(g-1)/\sin^2x\bigr),\\
  \mathcal{E}_n&=n(n/2+g),\quad
  n=0,1,2, \ldots,\quad g>0,\quad 0<x<\pi,\quad \eta(x)=\cos x, \\
  \phi_n(x)&=(\sin x)^gP_n^{(\beta,\beta)}(\cos x),
  \quad \beta\eqdef g-1/2,
\end{align}
in which $P_n^{(\alpha,\beta)}(\eta)$ is the Jacobi polynomial 
\eqref{defJac} and $P_n^{(\beta,\beta)}(\eta)$ is proportional to 
the Gegenbauer polynomial $C_n^{(\beta+1/2)}(\eta)$ \eqref{defGege}
\begin{equation}
  \frac{P_n^{(\beta,\beta)}(\eta)}{(\beta+1)_n}
  =\frac{C_n^{(\beta+1/2)}(\eta)}{(2\beta+1)_n}.
  \label{gegen}
\end{equation}
Hereafter we often use the Pochhammer symbol $(a)_n$, see \eqref{defPoch}.

It is straightforward to evaluate the Poisson brackets
\begin{equation}
  \{\mathcal{H},\cos x\}=p\sin x,\quad
  \{\mathcal{H},\{\mathcal{H}, \cos x\}\}=-\cos x\,\,2\mathcal{H}',\quad 
  \mathcal{H}'\eqdef \mathcal{H}+g^2/2,
\end{equation}
leading to the solution of the initial value problem:
\begin{equation}
  \cos x(t)=\cos x(0) \cos\bigl[t\sqrt{2\mathcal{H}'_0}\,\bigr] -p(0)\sin
  x(0)\frac{\sin\bigl[t\sqrt{2\mathcal{H}'_0}\,\bigr]}
  {\sqrt{2\mathcal{H}'_0}}.
\end{equation}
It is straightforward to verify $|\cos x(t)|<1$.
The corresponding quantum expressions are
\begin{align}
  \lbrack \mathcal{H},\cos x]&=i\sin x\, p+ \cos x/2, \\
  \lbrack \mathcal{H},[\mathcal{H},\cos x]\,]&=
  \cos x(2\mathcal{H}'-1/4)+[\mathcal{H},\cos x],\\
  \alpha_\pm(\mathcal{H})&=1/2\pm\sqrt{2\mathcal{H}'}.
\end{align}
The exact operator solution reads
\begin{align}
  e^{it\mathcal{H}}\cos x\,e^{-it\mathcal{H}}&=
  (i\sin x\, p+ \cos x/2)
  \frac{e^{i\alpha_+(\mathcal{H})t}-e^{i\alpha_-(\mathcal{H})t}}
  {2\sqrt{2\mathcal{H}'}}\n
  &\quad
  +\cos x\,\frac{-\alpha_-(\mathcal{H})e^{i\alpha_+(\mathcal{H})t}+
  \alpha_+(\mathcal{H})e^{i\alpha_-(\mathcal{H})t}}
  {2\sqrt{2\mathcal{H}'}}.
\end{align}
The annihilation and creation operators are
\begin{equation}
  a^{\prime(\pm)}=a^{(\pm)}2\sqrt{2\mathcal{H}'}
  =\pm i\sin x\, p+ \cos x\,\sqrt{2\mathcal{H}'}
  =\pm \sin x\frac{d}{dx}+\cos x\,\sqrt{2\mathcal{H}'}.
\end{equation}
It is now obvious that they ($a^{\prime(\pm)}$) are not hermitian 
conjugate to each other.
The square root sign is neatly removed when applied to the eigenvector 
$\phi_n$ as $2\mathcal{E}_n+g^2=(n+g)^2$:
\begin{align}
  a^{\prime(-)}\phi_n&=
  -\sin x\frac{d\phi_n}{dx}+(n+g)\cos x\,\phi_n=(n+\beta)\phi_{n-1},
  \label{suta'pn}\\
  a^{\prime(+)}\phi_n&=\phantom{-}\sin x\frac{d\phi_n}{dx}
  +(n+g)\cos x\,\phi_n
  =\frac{2(n+1)(n+2g)}{2n+2g+1}\phi_{n+1}.
 \label{suta'pn1}
\end{align}
This is a rule rather than exception as expected from the relations
between the neighbouring energy levels, \eqref{cond1} and \eqref{cond2}.
The right hand sides are the results of the application.
In particular, when acting on the ground state $\phi_0$, the annihilation
($a^{\prime(-)}$)
and creation ($a^{\prime(+)}$) operators are proportional to the
factorisation operators $\mathcal{A}$ and $\mathcal{A}^\dagger$ of the
Hamiltonian $\mathcal{H}=\mathcal{A}^\dagger\mathcal{A}/2$, respectively:
\begin{align}
  a^{\prime(-)}\phi_0&=\bigl(-\sin x\frac{d}{dx}+g\cos x\bigr)\phi_0
  =-\sin x\bigl(\frac{d}{dx}-g\cot x\bigr)\phi_0
  =\eta'(x)\mathcal{A}\phi_0=0,\n
  a^{\prime(+)}\phi_0&=\bigl(\sin x\frac{d}{dx}+g\cos x\bigr)\phi_0
  =-\sin x\bigl(-\frac{d}{dx}-g\cot x\bigr)\phi_0
  =\eta'(x)\mathcal{A}^{\dagger}\phi_0.
\end{align}
In a rough sense, the factor $\eta'(x)=-\sin x$, in the creation operator,
compensates the downward shift of the parameter ($g$) caused by
$\mathcal{A}^\dagger$.
Similar situations are encountered in all the other quantum systems.
In particular, for the systems with equi-spaced spectrum $\mathcal{E}_n=an$
($a$ : constant), the factorisation of  $a^{\prime(-)}$
and  $a^{\prime(+)}$ into $\mathcal{A}$ and $\mathcal{A}^\dagger$
is $n$-independent, \eqref{acfact}, \eqref{calfacta}, \eqref{calfactc},
\eqref{cdHacfac}. Their significance will be discussed in some detail
for the `discrete' quantum mechanics cases in Appendix {\bf B}.

The following interesting commutation relations ensue from
\eqref{suta'pn} and \eqref{suta'pn1}:
\begin{align}
  [\mathcal{H},a^{\prime(\pm)}]&=
  \pm\bigl(\sqrt{2\mathcal{H}'}\,a^{\prime(\pm)}
  +a^{\prime(\pm)}\sqrt{2\mathcal{H}'}\,\bigr)/2,
  \label{acal1}\\
  [a^{\prime(-)},a^{\prime(+)}]&=2\sqrt{2\mathcal{H}'},
  \label{acal2}\\
  a^{\prime(-)}a^{\prime(+)}+a^{\prime(+)}a^{\prime(-)}&=
  4\mathcal{H}+2g.
  \label{acal3}
\end{align}
The relation \eqref{acal3} could be accepted as a substitute
of the property (\romannumeral3) of the annihilation-creation operators 
discussed in page \pageref{acprop}.

By the similarity transformation in terms of the ground state 
wavefunction $\phi_0(x)= (\sin x)^g$, we obtain the so-called 
shift down and up operators for the Jacobi (Gegenbauer) polynomial 
($\beta=g-1/2$):
\begin{align}
  \mbox{down}&\ :\ \phantom{-}
  (1-\eta^2)\frac{d}{d\eta}P_n^{(\beta,\beta)}(\eta)
  +n \eta P_n^{(\beta,\beta)}(\eta) 
  =(n+\beta)P_{n-1}^{(\beta,\beta)}(\eta),
  \label{jacdown}\\
  \mbox{up}&\ :\ -(1-\eta^2)\frac{d}{d\eta}P_n^{(\beta,\beta)}(\eta)
  +(n+2g) \eta P_n^{(\beta,\beta)}(\eta) 
  =\frac{2(n+1)(n+2g)}{2n+2g+1}P_{n+1}^{(\beta,\beta)}(\eta).
\end{align}
As expected they are the same Jacobi polynomials of degree 
$n-1$ and $n+1$. It should be stressed that these shift down-up 
operators are naturally derived from our annihilation-creation 
operators without assuming the explicit form of the three term 
recursion relation.

\bigskip
The coherent state $\psi$ \eqref{cohedef} is
\begin{equation}
  \psi(x)=\phi_0(x)\sum_{n=0}^\infty(2\lambda)^n
  \frac{(\beta+3/2)_n}{(\beta+1)_n}P_n^{(\beta,\beta)}(\cos x)
  =\phi_0(x)\sum_{n=0}^\infty(2\lambda)^n\frac{(g+1)_n}{(2g)_n}
  C_n^{(g)}(\cos x),
\end{equation}
where we have used $a^{(-)}\phi_n=a^{\prime(-)}\phi_n/(2(n+g))$, 
\eqref{suta'pn} and \eqref{gegen}.
A generating function of the Gegenbauer polynomials ($\gamma$: arbitrary)
\cite{koeswart},
\begin{equation}
  \sum_{n=0}^\infty t^n\frac{(\gamma)_n}{(2g)_n}C_n^{(g)}(\eta)
  =(1-\eta t)^{-\gamma}\,
  {}_2F_1\Bigl(\genfrac{}{}{0pt}{}{\gamma/2,(\gamma+1)/2}{g+1/2}\Bigm|
   \frac{(\eta^2-1)t}{(1-\eta t)^2}\Bigr),
\end{equation}
gives a concise expression of the coherent state $\psi$:
\begin{eqnarray}
  \psi(x)=(1-2\lambda\cos x)^{-g-1}\,
  {}_2F_1\Bigl(\genfrac{}{}{0pt}{}{(g+1)/2,g/2+1}{g+1/2}\Bigm|
   \frac{-2\lambda\sin^2x}{(1-2\lambda\cos x)^2}\Bigr).
  \label{sutcoherent1}
\end{eqnarray}
Here ${}_2F_1$ is the hypergeometric function \eqref{defhypergeom}.
For the annihilation operator $a^{\prime(-)}$, the corresponding
coherent state $\psi'$ is 
\begin{equation}
  \psi'(x)=\phi_0(x)\sum_{n=0}^\infty\frac{\lambda^n}{(\beta+1)_n}
  P_n^{(\beta,\beta)}(\cos x)
  =\phi_0(x)\sum_{n=0}^\infty\frac{\lambda^n}{(2g)_n}C_n^{(g)}(\cos x).
\end{equation}
A generating function of the Gegenbauer polynomials \cite{koeswart},
\begin{equation}
  \sum_{n=0}^\infty\frac{t^n}{(2g)_n}C_n^{(g)}(\eta)
  =e^{\eta t}{}_0F_1\Bigl(\genfrac{}{}{0pt}{}{-}{g+1/2}\Bigm|
   \frac{(\eta^2-1)t^2}{4}\Bigr),
\end{equation}
gives a concise expression of the coherent state $\psi'$:
\begin{eqnarray}
  \psi'(x)=\Gamma(g+1/2)\,e^{\lambda\cos x}
  ({\lambda/2})^{1/2-g}\sqrt{\sin x}
  \,J_{g-1/2}(\lambda\sin x),
  \label{sutcoherent2}
\end{eqnarray}
in which $J_a(x)$ is the Bessel function \eqref{defBes}. 

\subsubsection{Deformed harmonic oscillator $\Rightarrow$
Meixner-Pollaczek polynomial}
\label{defosci}

The deformed harmonic oscillator is a simplest example of shape invariant
`discrete' quantum mechanics.
The Hamiltonian of `discrete' quantum mechanics studied in this paper
has the following form \cite{os4} (with some modification for
the Askey-Wilson case in  section \ref{aswil}):
\begin{equation}
  \mathcal{H}\eqdef\Bigl(\sqrt{V(x)}\,e^{\,p}\sqrt{V(x)^*}
  +\sqrt{V(x)^*}\,e^{-p}\sqrt{V(x)}
  -V(x)-V(x)^*\Bigr)\!\bigm/2.
  \label{H}
\end{equation}
The eigenvalue problem for $\mathcal{H}$, 
$\mathcal{H}\phi=\mathcal{E}\phi$ is a difference equation,
instead of a second order differential equation.
Let us define $S_{\pm}$, $T_{\pm}$ and $\mathcal{A}$ by
\begin{gather}
  S_+\eqdef e^{p/2}\sqrt{V(x)^*},\ \,
  S_-\eqdef e^{-p/2}\sqrt{V(x)},\ \,
  S_+^{\dagger}=\sqrt{V(x)}\,e^{p/2},\ \,
  S_-^{\dagger}=\sqrt{V(x)^*}\,e^{-p/2},
  \label{spmdef}\\
  T_+\eqdef S_+^{\dagger}S_+=\sqrt{V(x)}\,e^p\sqrt{V(x)^*},\quad
  T_-\eqdef S_-^{\dagger}S_-=\sqrt{V(x)^*}\,e^{-p}\sqrt{V(x)},
  \label{tpmdef}\\
  \mathcal{A}\eqdef i(S_+-S_-),\quad
  \mathcal{A}^{\dagger}=-i(S_+^{\dagger}-S_-^{\dagger}).
  \label{aaddef}
\end{gather}
Then the Hamiltonian is factorized
\begin{equation}
  \mathcal{H}=\bigl(T_++T_--V(x)-V(x)^*\bigr)/2
  =(S_+^{\dagger}-S_-^{\dagger})(S_+-S_-)/2
  =\mathcal{A}^{\dagger}\mathcal{A}/2.
\end{equation}

The potential function $V(x)$ of the deformed harmonic oscillator is
\begin{equation}
  V(x)=a+ix,\quad -\infty<x<\infty,\quad a>0.
\end{equation}
As shown in some detail in our previous paper \cite{os4},
it has an equi-spaced spectrum and the corresponding eigenfunctions
are a special case of the Meixner-Pollaczek polynomial
$P_n^{(a)}(x\,;\tfrac{\pi}{2})$ \eqref{defMP},
\begin{align}
  \mathcal{E}_n&=n,\quad n=0,1,2,\ldots,\\
  \phi_0(x)&=\sqrt{\Gamma(a+ix)\Gamma(a-ix)},\quad \eta(x)=x,
  \label{phi0MP}\\
  \phi_n(x)&=\phi_0(x)P_n(x),\quad
  P_n(x)\eqdef P_n^{(a)}(x\,;\tfrac{\pi}{2}),
  \label{phinMP}  
\end{align}
which could be considered as a deformation of the Hermite polynomial.

The Poisson bracket relations are
\begin{equation}
  \{\mathcal{H},x\}=-\sqrt{a^2+x^2}\sinh p,\qquad
  \{\mathcal{H},\{\mathcal{H},x\}\}=-x,
\end{equation}
leading to the harmonic oscillation,
\begin{equation}
  x(t)=x(0)\cos t+\sqrt{a^2+x^2(0)}\,\sinh p(0)\, \sin t,
\end{equation}
which endorses the naming of the deformed harmonic oscillator.
The corresponding quantum expressions are also simple:
\begin{align}
  [\mathcal{H},x]&=-i(T_+-T_-)/2,\qquad 
  [\mathcal{H},[\mathcal{H},x]\,]=x, \\
  e^{it\mathcal{H}}\,x\,e^{-it\mathcal{H}}&= x\,\cos t
  +i[\mathcal{H},x]\,\sin t= x\,\cos t+(T_+-T_-)/2\sin t.
\end{align}
The annihilation and creation operators are
\begin{eqnarray}
  a^{\prime(\pm)}=2a^{(\pm)}=x\pm[\mathcal{H},x]=x\mp i(T_+-T_-)/2,
\end{eqnarray}
which are hermitian conjugate to each other.
These operators were also introduced by Degasperis and Ruijsenaars
\cite{degruij} by a different reasoning from ours.
By similarity transformation  in terms of the ground state wavefunction
$\phi_0(x)= \sqrt{\Gamma(a+ix)\Gamma(a-ix)}$, we obtain
\begin{equation}
  \phi_0(x)^{-1}\ a^{\prime(\pm)}\ \phi_0(x)
  =x\mp i\bigl(V(x)e^p-V(x)^*e^{-p}\bigr)/2\,.
\end{equation}
The action of the annihilation creation operators on the eigenvectors
\begin{equation}
  a^{\prime(-)}\phi_n=(n+2a-1)\phi_{n-1},\quad
  a^{\prime(+)}\phi_n=(n+1)\phi_{n+1}
  \label{mxpoldown}
\end{equation}
is consistent with 
the three term recurrence relation of the Meixner-Pollaczek polynomial:
\begin{equation}
  (n+1)P_{n+1}^{(a)}(x\,;\tfrac{\pi}{2})-2xP_n^{(a)}(x\,;\tfrac{\pi}{2})
  +(n+2a-1)P_{n-1}^{(a)}(x\,;\tfrac{\pi}{2})=0.
\end{equation}
{}From these it is easy to verify the $\mathfrak{su}(1,1)$ commutation 
relations including the Hamiltonian $\mathcal{H}$:
\begin{equation}
  [\mathcal{H},a^{\prime(\pm)}]=\pm\, a^{\prime(\pm)},\quad 
  [a^{\prime(-)},a^{\prime(+)}]=2(\mathcal{H}+a).
\end{equation}
It is interesting to note that $a^{(\pm)}$ are factorised by the factors
of the Hamiltonian $\mathcal{H}=\mathcal{A}^\dagger \mathcal{A}/2$,
\begin{gather}
  4a^{(-)}=X^\dagger\mathcal{A},\quad
  4a^{(+)}=\mathcal{A}^\dagger X,
  \label{acfact}\\
  X\eqdef S_++S_-,\quad X^{\dagger}=S_+^{\dagger}+S_-^{\dagger}. 
  \label{XMP}
\end{gather}
These $X$ and $X^{\dagger}$ compensate the shift of the parameter $a$
caused by $\mathcal{A}^\dagger$ and $\mathcal{A}$.
See Appendix {\bf B} for more details.

\bigskip
The coherent state \eqref{cohedef}, \eqref{cohe}, is simply obtained
from the formula \eqref{mxpoldown} and
$a^{\prime(-)}=2a^{(-)}$:
\begin{equation}
  \psi(x)=\phi_0(x)\sum_{n=0}^\infty\frac{(2\lambda)^n}{(2a)_n}
  P_{n}^{(a)}(x\,;\tfrac{\pi}{2}).
\end{equation}
A generating function of the Meixner-Pollaczek polynomial \cite{koeswart},
\begin{equation}
  \sum_{n=0}^\infty\frac{t^n}{(2a)_n}P_{n}^{(a)}(x\,;\tfrac{\pi}{2})
  =e^{it}{}_1F_1\Bigl(\genfrac{}{}{0pt}{}{a+ix}{2a}\Bigm|
  -2it\Bigr),
\end{equation}
gives a concise expression of the coherent state $\psi$:
\begin{eqnarray}
  \psi(x)=\phi_0(x)\,
  e^{2i\lambda}\,{}_1F_1\Bigl(\genfrac{}{}{0pt}{}{a+ix}{2a}\Bigm|
   -4i\lambda\Bigr),
\end{eqnarray}
in which ${}_1F_1$ is the hypergeometric function \eqref{defhypergeom}. 

\subsubsection{Askey-Wilson polynomial}
\label{aswil}

The Askey-Wilson polynomial belongs to the so-called $q$-scheme of
hypergeometric polynomials \cite{koeswart}. It has four parameters
$a_1,a_2,a_3,a_4$ on top of $q$ ($0<q<1$), and is considered as a
three-parameter deformation of the Jacobi polynomial. As shown in our
previous papers
\cite{os3,os4}, it also describes the equilibrium positions of the
trigonometric  Ruijsenaars-Schneider systems based on the $BC$ root
system \cite{RSvD}.  Thus, as a dynamical system, it could be called a
deformed  P\"oschl-Teller potential or one body case of the
trigonometric $BC$ Ruijsenaars-Schneider systems.
The quantum-classical correspondence has some more subtlety than the
other `discrete' quantum mechanical systems treated in section \ref{dqm}
because of another `classical' limit $q\to1$.

The factorised Hamiltonian of the Askey-Wilson polynomial
has a bit different form from that of the Meixner-Pollaczek polynomial
\eqref{H}:
\begin{equation}
  \mathcal{H}\eqdef\Bigl(\sqrt{V(z)}\,q^{D}\!\sqrt{V(z)^*}
  +\sqrt{V(z)^*}\,q^{-D}\!\sqrt{V(z)}
  -V(z)-V(z)^*\Bigr)\!\bigm/2,
  \label{asH}
\end{equation}
with a potential function $V(z)$:
\begin{equation}
  V(z)=\frac{\prod_{j=1}^4(1-a_jz)}{(1-z^2)(1-qz^2)}\,,\quad 
  z=e^{ix},\quad 0<x<\pi,\quad
  D\eqdef z\frac{d}{dz}=-i\frac{d}{dx}=p.
  \label{askeypot}
\end{equation}
We assume $-1<a_1,a_2,a_3,a_4<1$ and $a_1a_2a_3a_4<q$.
This Hamiltonian is also factorised 
$\mathcal{H}=\mathcal{A}^{\dagger}\mathcal{A}/2$, where
$\mathcal{A}$ and $\mathcal{A}^{\dagger}$ are given in 
\eqref{spmdef}--\eqref{aaddef} with the replacement 
$V(x)\Rightarrow V(z)$, $e^{\pm p/2}\Rightarrow q^{\pm D/2}$, etc.
The eigenvalues and eigenfunctions are \cite{os4,os3}:
\begin{align}
  \mathcal{E}_n&=(q^{-n}-1)(1-a_1a_2a_3a_4q^{n-1})/2,\quad
  n=0,1,2,\ldots,\\
  \phi_0(x)&=\sqrt{
  \frac{(z^2\,;q)_{\infty}}{\prod_{j=1}^4(a_jz\,;q)_{\infty}}
  \frac{(z^{-2}\,;q)_{\infty}}{\prod_{j=1}^4(a_jz^{-1}\,;q)_{\infty}}}\,,
  \quad
  \eta(x)=\frac{z+z^{-1}}{2}=\cos x,
  \label{phi0AW}\\
  \phi_n(x)&=\phi_0(x)P_n(\cos x),\quad
  P_n(\eta)\eqdef p_n(\eta\,;a_1,a_2,a_3,a_4|q),
  \label{phinAW}
\end{align}
in which $p_n(\eta\,;a_1,a_2,a_3,a_4|q)$ is the Askey-Wilson polynomial
\eqref{defAW}.

The presence of the $q$-factor has only superficial effects at the
classical level with the Hamiltonian ($\gamma=\log q$):
\begin{equation}
  \mathcal{H}_c=\sqrt{V_c(z)V_c(z)^*}\cosh\gamma p
  -\bigl(V_c(z)+V_c(z)^*\bigr)/2,\quad 
  V_c(z)=\frac{\prod_{j=1}^4(1-a_jz)}{(1-z^2)^2},
\end{equation}
\begin{align}
  \{\mathcal{H}_c,\cos x\}&=\gamma
  \mbox{$\sqrt{\prod_{j=1}^4(1-a_j z)\prod_{j=1}^4(1-a_j/z)}$}
  \!\bigm/(4\sin x)\ \sinh\gamma p,\\
  \{\mathcal{H}_c,\{\mathcal{H}_c,\cos x\}\}&=-\cos x\,
  R_0(\mathcal{H}_c)-R_{-1}(\mathcal{H}_c),\\
  R_0(\mathcal{H}_c)&=\gamma^2(\mathcal{H}_c^2+c_1\mathcal{H}_c+c_2), 
  \quad R_{-1}(\mathcal{H}_c)=-\gamma^2(c_3\mathcal{H}_c+c_4),
\end{align}
with  coefficients $c_1$,..,$c_4$:
\begin{equation}
  c_1=1+b_4,\quad c_2=(1-b_4)^2/4,\quad c_3=(b_1+b_3)/4,\quad
  c_4=(1-b_4)(b_1-b_3)/8.
\end{equation}
Here we use the abbreviation
\begin{equation}
   b_1\eqdef\sum_{1\leq j\leq 4}a_j\,,\quad
   b_3\eqdef\!\!\!\sum_{1\leq j<k<l\leq 4}a_ja_ka_l\,,\quad
   b_4\eqdef\prod_{j=1}^4a_j\,.
\end{equation}

The corresponding quantum expressions are
\begin{align}
  [\mathcal{H},\cos x]&=(q^{-1}-1)\bigl(z^{-1}(1-qz^2)T_+
  +z(1-qz^{-2})T_-\bigr)/4,\\
  \ [\mathcal{H},[\mathcal{H},\cos x]\,]&=\cos x\,R_0(\mathcal{H})+
  [\mathcal{H},\cos x]R_1(\mathcal{H})+R_{-1}(\mathcal{H}),\\
  R_0(\mathcal{H})&=q(q^{-1}-1)^2\Bigl((\mathcal{H}')^2
  -(1+q^{-1})^2b_4/4\Bigr)\,,\\
  R_1(\mathcal{H})&=q(q^{-1}-1)^2\,\mathcal{H}', \qquad\quad 
  \mathcal{H}'\eqdef\mathcal{H}+(1+q^{-1}b_4)/2,\\
  R_{-1}(\mathcal{H})&=-q(q^{-1}-1)^2\Bigl((b_1+q^{-1}b_3)\mathcal{H}/4
   +(1-q^{-2}b_4)(b_1-b_3)/8\Bigr).
\end{align}
The two frequencies are:
\begin{equation}
  \alpha_{\pm}(\mathcal{H})=(q^{-1}-1)\Bigl((1-q)\mathcal{H}'
  \pm(1+q)\sqrt{(\mathcal{H}')^2-q^{-1}b_4}\,\Bigr)\!\bigm/2,
\end{equation}
in which
\begin{equation}
  \mathcal{H}'\phi_n=(q^{-n}+b_4q^{n-1})/2\ \phi_n,\quad
  \bigl((\mathcal{H}')^2-q^{-1}b_4\bigr)\phi_n
  =(q^{-n}-b_4q^{n-1})^2/4\ \phi_n.
\end{equation}
The annihilation-creation operators are:
\begin{align}
  a^{(\pm)}&=\Bigl(
  \pm(q^{-1}-1)\bigl(z^{-1}(1-qz^2)T_++z(1-qz^{-2})T_-\bigr)/4\n
  &\qquad \mp\cos x\,\alpha_{\mp}(\mathcal{H})
  \pm R_{-1}(\mathcal{H})\alpha_{\pm}(\mathcal{H})^{-1}
  \Bigr)\bigm/\bigl(\alpha_+(\mathcal{H})-\alpha_-(\mathcal{H})\bigr)\,.
\end{align}
Their effects on the eigenvectors are:
\begin{align}
  a^{(-)}\phi_n&
  =\frac{(1-q^n)\prod_{1\leq j<k\leq 4}(1-a_ja_kq^{n-1})}
  {2(1-b_4q^{2n-2})(1-b_4q^{2n-1})}\,\phi_{n-1},\\
  a^{(+)}\phi_n&=\frac{1-b_4q^{n-1}}
  {2(1-b_4q^{2n-1})(1-b_4q^{2n})}\,\phi_{n+1},
\end{align}
which are consistent with the three term recurrence relation of
the Askey-Wilson polynomial.
The `annihilation-creation' operators on the polynomial $P_n(\cos x)$
read
\begin{align}
  &\phi_0(x)^{-1}a^{(\pm)}\phi_0(x)\cdot P_n(\cos x)\n
  =\,&\frac{1}{\mathcal{E}_{n+1}-\mathcal{E}_{n-1}}\Bigl(
  \pm(q^{-1}-1)\Bigl(z^{-1}(1-qz^2)V(z)\,q^{D}
  +z(1-qz^{-2})V(z)^*q^{-D}\Bigr)\!\bigm/4\n
  &\qquad\qquad\qquad
  \pm(\mathcal{E}_n-\mathcal{E}_{n\mp 1})\cos x
  \pm\frac{R_{-1}(\mathcal{E}_n)}{\mathcal{E}_{n\pm 1}-\mathcal{E}_n}
  \Bigr)P_n(\cos x).
\end{align}
The coherent state is
\begin{equation}
  \psi(x)=\phi_0(x)\sum_{n=0}^{\infty}
  \frac{(2\lambda)^n}{(q\,;q)_n}\,
  \frac{(a_1a_2a_3a_4\,;q)_{2n}}{\prod_{1\leq j<k\leq 4}(a_ja_k\,;q)_n}\,
  P_n(\cos x)\,.
\end{equation}
We are not aware if a concise summation formula exists or not.

\section{Various results}
\label{3rd}
\setcounter{equation}{0}

In this section we will briefly present many interesting results 
on the annihilation-creation operators, their algebraic properties 
and coherent states, etc, for various exactly solvable quantum 
mechanical systems, including the `discrete' quantum mechanical 
systems. 
All these solvable systems share {\em shape invariance\/}
\cite{genden,susyqm,os4} which is a purely quantum mechanical notion
that guarantees quantum solvability.  But that property plays no active
role in the present theory. On the other hand, the `sinusoidal motion'
exists at the classical and quantum levels.
In Appendix {\bf A} we will show 
that a system realising the exact sinusoidal motion is quite limited 
and that all of them belong to the known shape invariant systems.
In other words there are shape invariant systems that do not have 
sinusoidal motion.

\subsection{Ordinary quantum mechanical systems}
\label{oqm}

\subsubsection{$x^2+1/x^2$ potential}
\label{calo}

When a centrifugal barrier (a $1/x^2$ potential) is added, 
the harmonic oscillator keeps its exact solvability, but 
the particle is restricted to a half line, either $x>0$ or
$x<0$. This is the one-body case of the well-known Calogero model
\cite{calsut,kps}.
The eigenfunctions are described by the Laguerre polynomial
and the annihilation-creation operators within the 
$\mathfrak{su}(1,1)$ scheme are well-known \cite{nieto,agch}.
Our unified theory predicts these operators naturally. The coherent and
squeezed states in the $\mathfrak{su}(1,1)$ were already known 
\cite{agch}.
Its Hamiltonian, the eigenvalues and the eigenfunctions are:
\begin{align}
  \mathcal{H}&\eqdef(p+ix-ig/x)(p-ix+ig/x)/2,\quad 0<x<\infty,
  \quad g>0,\\
  \mathcal{E}_n&=2n,\quad
  n=0,1,2,\ldots,\quad \eta(x)=x^2,\\
  \phi_n(x)&=\phi_n(x\,;g)=e^{-x^2/2}x^gL_n^{(\beta)}(x^2),
  \quad \beta\eqdef g-1/2,
\end{align}
in which $L_n^{(\beta)}(\eta)$ is the Laguerre polynomial \eqref{defLag}.

The Poisson bracket relations are simple
\begin{equation}
  \{\mathcal{H},x^2\}=-2px,\quad \{\mathcal{H},\{\mathcal{H},x^2\}\}
  =-4(x^2-\mathcal{H}-g),
\end{equation}
leading to the simple sinusoidal motion
\begin{equation}
  x^2(t)=x^2(0)\,\cos 2t+(1-\cos2t)(\mathcal{H}_0+g)+p(0)x(0)\sin2t.
\end{equation}
It is straightforward to verify $x^2(t)>0$. 
The quantum theory is almost the same as the classical one:
\begin{align}
  \lbrack \mathcal{H},x^2]&=-i(xp+px),\quad
  \lbrack \mathcal{H},[\mathcal{H},x^2]\,]=4(x^2-\mathcal{H}'),\quad 
  \mathcal{H}'\eqdef\mathcal{H}+g+1/2, \\
  e^{it\mathcal{H}}x^2\,e^{-it\mathcal{H}}&=
  x^2\,\cos2t+(1-\cos2t)\mathcal{H}'+(xp+px)/2\,\sin2t,
\end{align}
which leads to the following annihilation and creation operators:
\begin{equation}
  a^{(\pm)}=(x^2-\mathcal{H}')/2\mp i(xp+px)/4
  =\Bigl(\bigl(\frac{d}{dx}\mp x\bigr)^2-\frac{g(g-1)}{x^2}\Bigr)\!\bigm/4.
  \label{ac1/x^2}
\end{equation}
The action of these operators are ($\beta=g-1/2$)
\begin{equation}
  a^{(-)}\phi_n=-(n+\beta)\phi_{n-1},\quad
  a^{(+)}\phi_n=-(n+1)\phi_{n+1},
\end{equation}
which are consistent with the three term recurrence relation of 
the Laguerre polynomial
\begin{equation}
  (n+1)L_{n+1}^{(\beta)}(\eta)+(\eta-2n-\beta-1)L_{n}^{(\beta)}(\eta)
  +(n+\beta)L_{n-1}^{(\beta)}(\eta)=0.
\end{equation}
{}From these follow the $\mathfrak{su}(1,1)$ relations
\begin{equation}
  [\mathcal{H},a^{(\pm)}]=\pm 2a^{(\pm)},\quad 
  [a^{(-)},a^{(+)}]=\mathcal{H}'=\mathcal{H}+g+1/2.
\end{equation}
The coherent state (AOCS) \eqref{cohe} is obtained simply as
\begin{equation}
  \psi(x)=\sum_{n=0}^\infty\frac{(-\lambda)^n}{(\beta+1)_n}\,\phi_n(x)
  =\phi_0(x)\frac{e^{-\lambda}\Gamma(\beta+1)}{(-x^2\lambda)^{\beta/2}}
  J_\beta(2x\sqrt{-\lambda}),
\end{equation}
in which a generating function of the Laguerre polynomial \cite{koeswart}
\begin{equation}
  \sum_{n=0}^\infty\frac{t^n}{(\alpha+1)_n}L_n^{(\alpha)}(x)
  =e^t\,
  {}_0F_1\Bigl(\genfrac{}{}{0pt}{}{-}{\alpha+1}\Bigm|-xt\Bigr)
\end{equation}
and \eqref{defBes} are used.

The annihilation and creation operators \eqref{ac1/x^2} 
are factorised by the factors of the Hamiltonian 
$\mathcal{H}=\mathcal{A}^\dagger \mathcal{A}/2$:
\begin{alignat}{2}
  4a^{(-)}&=
  \Bigl(\frac{d}{dx}+x+\frac{g}{x}\Bigr)\cdot\mathcal{A},
  & \mathcal{A}&=i(p-ix+ig/x),
  \label{calfacta}\\
  4a^{(+)}&= 
  \mathcal{A}^\dagger \cdot\Bigl(-\frac{d}{dx}+x+\frac{g}{x}\Bigr), 
  &\quad \mathcal{A}^\dagger&=-i(p+ix-ig/x).
  \label{calfactc}
\end{alignat}
This is a degenerate case of \eqref{cdHacfac} and 
the action of the other factors is to compensate  the parameter shifts
caused by the operators $\mathcal{A}^\dagger$ and $\mathcal{A}$:
\begin{align}
  \Bigl(-\frac{d}{dx}+x+\frac{g}{x}\Bigr)\phi_n(x\,;g)
  &=2\,\phi_n(x\,;g+1),\\
  \Bigl(\frac{d}{dx}+x+\frac{g}{x}\Bigr)\phi_n(x\,;g+1)
  &=(2n+2g+1)\,\phi_n(x\,;g).
\end{align}

\subsubsection{P\"oschl-Teller potential}
\label{poshtell}

The P\"oschl-Teller potential has two parameters $g$ and $h$ and its 
eigenfunctions are related to the Jacobi polynomial. It is  the one 
body case of the $BC$ type Sutherland systems \cite{calsut,kps}.
Its Hamiltonian, the eigenvalues and the eigenfunctions are:
\begin{align}
  \mathcal{H}&\eqdef(p-ig\cot x+ih\tan x)(p+ig\cot x-ih\tan x)/2,
  \quad 0< x<\pi/2, \\
  \mathcal{E}_n&=2n(n+g+h),\quad n=0,1,2, \ldots,\quad g,h>0,
  \quad \eta(x)=\cos 2x,\\
  \phi_n(x)&=(\sin x)^g(\cos x)^hP_n^{(\alpha,\beta)}(\cos 2x),
  \quad \alpha\eqdef g-1/2,\quad \beta\eqdef h-1/2,
\end{align}
in which $P_n^{(\alpha,\beta)}(\eta)$ is the Jacobi polynomial 
\eqref{defJac}.
It is straightforward to evaluate the Poisson brackets
\begin{align}
  \{\mathcal{H},\cos 2x\}&=2p\sin 2x,\\
  \{\mathcal{H},\{\mathcal{H}, \cos 2x\}\}
  &=-\cos 2x\, 8\mathcal{H}'-4(g^2-h^2),\quad 
  \mathcal{H}'\eqdef \mathcal{H}+(g+h)^2/2,
\end{align}
leading to the solution of the initial value problem:
\begin{align}
  \cos 2x(t)&=\Bigl(\cos 2x(0)+\frac{g^2-h^2}{2\mathcal{H}'_0}\Bigr)
  \cos\bigl[2t\sqrt{2\mathcal{H}'_0}\,\bigr]\n
  &\qquad\qquad\ -p(0)\sin 2x(0)
  \frac{\sin\bigl[2t\sqrt{2\mathcal{H}'_0}\,\bigr]}
  {\sqrt{2\mathcal{H}'_0}}-\frac{g^2-h^2}{2\mathcal{H}'_0}\,.
\end{align}
Note that $|\cos 2x(t)|<1$ is satisfied.
The corresponding quantum expressions are
\begin{align}
  \lbrack \mathcal{H},\cos 2x]&=2( i\sin 2x\,\, p+ \cos 2x),\\
  \lbrack \mathcal{H},[\mathcal{H},\cos 2x]\,]&=
  \cos 2x\,(8\mathcal{H}'-4)+4[\mathcal{H},\cos 2x]
  +4(\alpha^2-\beta^2),\\
  \alpha_\pm(\mathcal{H})&=2\pm 2\sqrt{2\mathcal{H}'}.
\end{align}
The exact operator solution reads
\begin{align}
  e^{it\mathcal{H}}\cos 2x\,e^{-it\mathcal{H}}&=
  (i\sin 2x\,\,p+ \cos 2x)
  \frac{e^{i\alpha_+(\mathcal{H})t}-e^{i\alpha_-(\mathcal{H})t}}
  {2\sqrt{2\mathcal{H}'}}
  -\frac{\alpha^2-\beta^2}{2\mathcal{H}'-1}\nonumber\\
  &\quad+\bigl(\cos 2x\,(2\mathcal{H}'-1)+\alpha^2-\beta^2\bigr)
  \frac{1}{\sqrt{2\mathcal{H}'}}
  \Bigl(\frac{e^{i\alpha_+(\mathcal{H})t}}{\alpha_+(\mathcal{H})}
  -\frac{e^{i\alpha_-(\mathcal{H})t}}{\alpha_-(\mathcal{H})}\Bigr).
\end{align}

The annihilation and creation operators are
\begin{equation}
  a^{\prime(\pm)}/2=a^{(\pm)}2\sqrt{2\mathcal{H}'}
  =\pm \sin 2x\frac{d}{dx}+\cos 2x\,\sqrt{2\mathcal{H}'}
  +\frac{\alpha^2-\beta^2}{\sqrt{2\mathcal{H}'}\pm 1}\,.
\end{equation}
It is now obvious that they ($a^{\prime(\pm)}$) are not hermitian 
conjugate to each other.
When applied to the eigenvector $\phi_n$ as 
$2\mathcal{E}_n+(g+h)^2=(2n+g+h)^2$, we obtain:
\begin{align}
  a^{\prime(-)}/2\,\phi_n
  &=-\sin 2x\frac{d\phi_n}{dx}+(2n+g+h)\cos 2x\,\phi_n
  +\frac{\alpha^2-\beta^2}{2n+\alpha+\beta}\,\phi_n\n
  &=\frac{4(n+\alpha)(n+\beta)}{2n+\alpha+\beta}\,\phi_{n-1},
  \label{PTdown}\\
  a^{\prime(+)}/2\,\phi_n
  &=\phantom{-}\sin 2x\frac{d\phi_n}{dx}+(2n+g+h)\cos 2x\,\phi_n
  +\frac{\alpha^2-\beta^2}{2n+\alpha+\beta+2}\,\phi_n\n
  &=\frac{4(n+1)(n+\alpha+\beta+1)}{2n+\alpha+\beta+2}\,\phi_{n+1}.
\end{align}
The right hand sides are the results of the application.
{}From \eqref{PTdown} and $a^{\prime(-)}\phi_n=4(2n+g+h)a^{(-)}\phi_n$, 
the coherent state $\psi$ and $\psi'$ are
\begin{align}
  \psi(x)&=\phi_0(x)\sum_{n=0}^\infty(\lambda/2)^n
  \frac{(\alpha+\beta+2)_{2n}}{(\alpha+1)_n(\beta+1)_n}\,
  P_n^{(\alpha,\beta)}(\cos 2x),\\
  \psi'(x)&=\phi_0(x)\sum_{n=0}^\infty(\lambda/4)^n
  \frac{(\frac{\alpha+\beta}{2}+1)_n}{(\alpha+1)_n(\beta+1)_n}\,
  P_n^{(\alpha,\beta)}(\cos 2x).
\end{align}
We are not aware if concise summation formulas exist or not.

\subsubsection{Soliton potential, or the symmetric Rosen-Morse
potential}
\label{soliton}

As is well-known $-{g(g+1)/{\cosh^2x}}$ potential is
{\em reflectionless} for integer coupling constant $g$,
corresponding to the KdV soliton. It has a finite number $1+[g]'$
(the greatest integer not equal or exceeding $g$) of bound states:
\begin{align}
  \mathcal{H}&\eqdef(p+ig\tanh x)(p-ig\tanh x)/2, \quad 
  -\infty< x<\infty,\quad g>0,\\
  \mathcal{E}_n&=n(-n/2+g),\quad n=0,1,\ldots,[g]',\quad\eta(x)=\sinh x,\\
  \phi_n(x)&=i^{-n}(\cosh x)^{-g}P_n^{(\beta,\beta)}(i\sinh x),\quad
  \beta\eqdef -g-1/2.
\end{align}
These eigenfunctions are real due to the parity
$P^{(\alpha,\beta)}_n(-x)=(-1)^nP^{(\beta,\alpha)}_n(x)$.
The Poisson brackets are
\begin{equation}
  \{\mathcal{H},\sinh x\}=-p\cosh x,\quad
  \{\mathcal{H},\{\mathcal{H}, \sinh x\}\}=\sinh x\,\,2\mathcal{H}',\quad 
  \mathcal{H}'\eqdef \mathcal{H}-g^2/2,
\end{equation}
leading to the solution of the initial value problem:
\begin{equation}
  \sinh x(t)=\sinh x(0)\cos\bigl[t\sqrt{-2\mathcal{H}'_0}\,\bigr]
  +p(0)\cosh x(0)
  \frac{\sin\bigl[t\sqrt{-2\mathcal{H}'_0}\,\bigr]}
  {\sqrt{-2\mathcal{H}'_0}}.
\end{equation}
It describes sinusoidal motion for bound states $\mathcal{H}_0'<0$ only.
But the above expression is valid for the unbound motion $\mathcal{H}_0'>0$,
too.

The corresponding quantum expressions are
\begin{align}
  \lbrack \mathcal{H},\sinh x]&= -i\cosh x\,\,p- \sinh x/2, \\
  \lbrack \mathcal{H},[\mathcal{H},\sinh x]\,]&=
  -\sinh x(2\mathcal{H}'+1/4)-[\mathcal{H},\sinh x],\\
  \alpha_\pm(\mathcal{H})&=-1/2\pm\sqrt{-2\mathcal{H}'}.
\end{align}
The exact operator solution reads
\begin{align}
  e^{it\mathcal{H}}\sinh x\,e^{-it\mathcal{H}}&=
  (-i\cosh x\,\,p-\sinh x/2)\,
  \frac{e^{i\alpha_+(\mathcal{H})t}-e^{i\alpha_-(\mathcal{H})t}}
  {2\sqrt{-2\mathcal{H}'}}\n
  &\quad-\sinh x\,\frac{2\mathcal{H}'+1/4}{2\sqrt{-2\mathcal{H}'}}
  \Bigl(\frac{e^{i\alpha_+(\mathcal{H})t}}{\alpha_+(\mathcal{H})}
  -\frac{e^{i\alpha_-(\mathcal{H})t}}{\alpha_-(\mathcal{H})}\Bigr).
\end{align}
The annihilation and creation operators are
\begin{equation}
  a^{\prime(\pm)}=a^{(\pm)}2\sqrt{-2\mathcal{H}'}
  =\mp\cosh x\frac{d}{dx}+\sinh x\,\sqrt{-2\mathcal{H}'}.
  \label{solannihi}
\end{equation}
When applied to the eigenvector $\phi_n$, we obtain as 
$2\mathcal{E}_n-g^2=-(g-n)^2$:
\begin{align}
  a^{\prime(-)}\phi_n&=
  \phantom{-}\cosh x\frac{d\phi_n}{dx}+(g-n)\sinh x\,\phi_n
  =(n+\beta)\phi_{n-1},\\
  a^{\prime(+)}\phi_n&=
  -\cosh x\frac{d\phi_n}{dx}+(g-n)\sinh x\,\phi_n
  =-\frac{(n+1)(n+2\beta +1)}{n+\beta+1}\phi_{n+1}.
\end{align}
We obtain the following interesting commutation relations:
\begin{align}
  [\mathcal{H},a^{\prime(\pm)}]&=
  \pm\bigl(\sqrt{-2\mathcal{H}'}\,a^{\prime(\pm)}
  +a^{\prime(\pm)}\sqrt{-2\mathcal{H}'}\,\bigr)/2,\\
  [a^{\prime(-)},a^{\prime(+)}]&=2\sqrt{-2\mathcal{H}'},
\qquad a^{\prime(-)}a^{\prime(+)}+a^{\prime(+)}a^{\prime(-)}
=4\mathcal{H}+2g,
\end{align}
which look very similar to those for the $1/\sin^2x$ potential
\eqref{acal1}--\eqref{acal3}.
In contrast to the $1/\sin^2x$ case, the present case has only
finite dimensional representation, $n=0,1,\ldots,[g]'$, that is
from the ground state to the highest level.
There is no coherent state as the eigenvector of the annihilation
operator \eqref{solannihi}.

\subsubsection{Morse potential}
\label{morse}

This is another well-known example of exactly solvable potential 
with a finite number of bound states \cite{susyqm}:
\begin{align}
  \mathcal{H}&\eqdef(p+i\mu\,e^x -ig)(p-i\mu\,e^x+ig)/2,
  \quad -\infty< x<\infty,\quad \mu,g>0,\\
  \mathcal{E}_n&=n(-n/2+g),\quad n=0,1, \ldots,[g]',\quad \eta(x)=e^{-x},\\
  \phi_n(x)&=e^{-\mu\,e^x+gx} e^{-nx}L_n^{(2g-2n)}(2\mu e^{x}).
\end{align}
The Poisson brackets are
\begin{equation}
  \{\mathcal{H},e^{-x}\}=p\,e^{-x},\quad
  \{\mathcal{H},\{\mathcal{H}, e^{-x}\}\}= e^{-x}2
  \mathcal{H}'+\mu g,\quad \mathcal{H}'\eqdef \mathcal{H}-g^2/2,
\end{equation}
leading to the solution of the initial value problem:
\begin{equation}
  e^{-x(t)}=\Bigl(e^{-x(0)}+\frac{\mu g}{2\mathcal{H}_0'}\Bigr)
  \cos\bigl[t\sqrt{-2\mathcal{H}'_0}\,\bigr]-p(0)e^{-x(0)}
  \frac{\sin\bigl[t\sqrt{-2\mathcal{H}'_0}\,\bigr]}
  {\sqrt{-2\mathcal{H}'_0}}-\frac{\mu g}{2\mathcal{H}_0'}.
\end{equation}
It describes sinusoidal motion for bound states $\mathcal{H}_0'<0$ only.
But the above expression is valid for the unbound motion $\mathcal{H}_0'>0$,
too.
It is easy to verify $e^{-x(t)}>0$.

The corresponding quantum expressions are
\begin{align}
  \lbrack \mathcal{H},e^{-x}]&= i\,e^{-x}p- e^{-x}/2, \\
  \lbrack \mathcal{H},[\mathcal{H},e^{-x}]\,]&=
  - e^{-x} (2\mathcal{H}'+1/4)-[\mathcal{H},e^{-x}] -\mu(g+1/2),\\
  \alpha_\pm(\mathcal{H})&=-1/2\pm\sqrt{-2\mathcal{H}'}.
\end{align}
The exact operator solution reads
\begin{align}
  e^{it\mathcal{H}}e^{-x}\,e^{-it\mathcal{H}}&=
  \bigl(i\,e^{-x}p- e^{-x}/2\bigr)
  \frac{e^{i\alpha_+(\mathcal{H})t}-e^{i\alpha_-(\mathcal{H})t}}
  {2\sqrt{-2\mathcal{H}'}}
  -\frac{\mu(g+1/2)}{2\mathcal{H}'+1/4}\n
  &\quad -\bigl(e^{-x}(2\mathcal{H}'+1/4)+\mu(g+1/2)\bigr)
  \frac{1}{2\sqrt{-2\mathcal{H}'}}
  \Bigl(\frac{e^{i\alpha_+(\mathcal{H})t}}{\alpha_+(\mathcal{H})}
   -\frac{e^{i\alpha_-(\mathcal{H})t}}{\alpha_-(\mathcal{H})}\Bigr).
\end{align}
The annihilation and creation operators are
\begin{equation}
  a^{\prime(\pm)}=a^{(\pm)}2\sqrt{-2\mathcal{H}'}
  =\pm e^{-x}\frac{d}{dx}+e^{-x}\sqrt{-2\mathcal{H}'}
  -\frac{\mu(2g+1)}{2\sqrt{-2\mathcal{H}'}\mp 1}\,.
\end{equation}
When applied to the eigenvector $\phi_n$, we obtain as 
$2\mathcal{E}_n-g^2=-(g-n)^2$:
\begin{align}
  a^{\prime(-)}\phi_n&=
  -e^{-x}\frac{d\phi_n}{dx}+(g-n)e^{-x}\phi_n
  -\frac{\mu(2g+1)}{2(g-n)+1}\phi_n
  =\frac{4\mu^2}{2(g-n)+1}\phi_{n-1},\\
  a^{\prime(+)}\phi_n&=
  \phantom{-}e^{-x}\frac{d\phi_n}{dx}+(g-n)e^{-x}\phi_n
  -\frac{\mu(2g+1)}{2(g-n)-1}\phi_n
  =\frac{(n+1)(2g-n)}{2(g-n)-1}\phi_{n+1}.
\end{align}

\subsection{`Discrete' quantum mechanical systems}
\label{dqm}

For specifying the dynamical systems belonging to the `discrete'
quantum mechanics \cite{os4,os3}, we use the name of the polynomial
eigenfunctions for want of universally accepted naming.
The factorised Hamiltonian is given by \eqref{H}.

\subsubsection{Continuous Hahn polynomial (special case)}
\label{chahn}

The factorised Hamiltonian of the continuous Hahn polynomial
(special case) has a potential function $V$ depending on two parameters:
\begin{equation}
  V(x)=(a_1+ix)(a_2+ix),\quad -\infty<x<\infty,\quad a_1,a_2>0.
  \label{conthahnpot}
\end{equation}
The eigenvalues and eigenfunctions are:
\begin{align}
  \mathcal{E}_n&=n(n+2a_1+2a_2-1)/2,\quad n=0,1,2,\ldots,\\
  \phi_0(x)&=\sqrt{{\textstyle\prod_{j=1}^2}\Gamma(a_j+ix)\Gamma(a_j-ix)},
  \quad \eta(x)=x,
  \label{phi0cH}\\
  \phi_n(x)&=\phi_0(x)P_n(x), \quad P_n(x)\eqdef p_n(x\,;a_1,a_2,a_1,a_2),
  \label{phincH}
\end{align}
in which $p_n(x\,;a_1,a_2,a_1,a_2)$ is a special case of the 
continuous Hahn polynomial \eqref{defcH}.
This is a two parameter deformation of the Hermite polynomial.
Thus this dynamical system is a deformed oscillator. 
The classical solution shows this fact clearly:
\begin{align}
  \{\mathcal{H},x\}&=-\sqrt{(a_1^2+x^2)(a_2^2+x^2)}\sinh p,\quad
  \{\mathcal{H},\{\mathcal{H},x\}\}
  =-x\bigl(2\mathcal{H}+(a_1+a_2)^2\bigr),\\[3pt]
  x(t)&=x(0)\cos\bigl[t\sqrt{2\mathcal{H}_0+(a_1+a_2)^2}\,\bigr]\n
  &\quad +\sqrt{(a_1^2+x(0)^2)(a_2^2+x(0)^2)}\,\sinh p(0)\,
  \frac{\sin\bigl[t\sqrt{2\mathcal{H}_0+(a_1+a_2)^2}\,\bigr]}
  {\sqrt{2\mathcal{H}_0+(a_1+a_2)^2}}.
\end{align}

The corresponding quantum solution is also simple:
\begin{align}
  [\mathcal{H},x]&=-i(T_+-T_-)/2,\\
  [\mathcal{H},[\mathcal{H},x]\,]&=
  x(2\mathcal{H}'-1/4)+[\mathcal{H},x],\quad
  2\mathcal{H}'\eqdef 2\mathcal{H}+(a_1+a_2-1/2)^2,\\
  e^{it\mathcal{H}}\,x\,e^{-it\mathcal{H}}&=
  [\mathcal{H},x]
  \frac{e^{i\alpha_+(\mathcal{H})t}-e^{i\alpha_-(\mathcal{H})t}}
  {2\sqrt{2\mathcal{H}'}}
  +x\,\frac{-\alpha_-(\mathcal{H})e^{i\alpha_+(\mathcal{H})t}
  +\alpha_+(\mathcal{H})e^{i\alpha_-(\mathcal{H})t}}
  {2\sqrt{2\mathcal{H}'}},\\
  \alpha_\pm(\mathcal{H})&=1/2\pm\sqrt{2\mathcal{H}'}.
\end{align}
The annihilation and creation operators are:
\begin{equation}
  a^{\prime(\pm)}=a^{(\pm)}2\sqrt{2\mathcal{H}'}
  =\pm [\mathcal{H},x]\mp x\,\alpha_{\mp}(\mathcal{H})
  =\mp i(T_+-T_-)/2+x(\sqrt{2\mathcal{H}'}\mp 1/2).
\end{equation}
When applied to the eigenvector $\phi_n$, we obtain as 
$2\mathcal{E}_n+(a_1+a_2-1/2)^2=(n+a_1+a_2-1/2)^2$:
\begin{align}
  2a^{\prime(-)}\phi_n&=
  \phantom{-}i(T_+-T_-)\phi_n+2x(n+a_1+a_2)\phi_n\n
  &=(n+a_1+a_2-1)(n+2a_1-1)(n+2a_2-1)\phi_{n-1},\\
  2a^{\prime(+)}\phi_n&=
  -i(T_+-T_-)\phi_n+2x(n+a_1+a_2-1)\phi_n\n
  &=\frac{(n+1)(n+2a_1+2a_2-1)}{n+a_1+a_2}\,\phi_{n+1}.
\end{align}
The similarity transformed operators act as
\begin{eqnarray}
  \phi_0(x)^{-1}a^{\prime(\pm)}\phi_0(x)\!\cdot\!P_n(x)
  =\Bigl(x(n+a_1+a_2-\tfrac12\mp\tfrac12)
  \mp\tfrac{i}{2}\bigl(V(x)e^p-V(x)^*e^{-p}\bigr)\Bigr)P_n(x).
\end{eqnarray}
The coherent state $\psi$ and $\psi'$ are
\begin{align}
  \psi(x)&=\phi_0(x)\sum_{n=0}^\infty
  \frac{\lambda^n(2a_1+2a_2)_{2n}}{(2a_1)_n(2a_2)_n(a_1+a_2)_n^2}\,P_n(x),\\
  \psi'(x)&=\phi_0(x)\sum_{n=0}^\infty
  \frac{(2\lambda)^n}{(2a_1)_n(2a_2)_n(a_1+a_2)_n}\,P_n(x).
\end{align}
We do not know if these sums have  concise expressions or not.

\subsubsection{Continuous dual Hahn polynomial}
\label{cdhahn}

The continuous dual Hahn polynomial has three parameters 
($a_1,a_2,a_3$) and is considered as a two parameter deformation of 
the Laguerre polynomial $L_n^{(\alpha)}$.
The factorised Hamiltonian of the continuous dual Hahn polynomial
has a potential function $V$:
\begin{equation}
  V(x)=\frac{\prod_{j=1}^3(a_j+ix)}{2ix(2ix+1)},\quad 
  0<x<\infty,\quad a_1,a_2,a_3>0.
  \label{dualconthahnpot}
\end{equation}
As a dynamical system this is  a deformed Calogero model, 
or a deformed $x^2+1/x^2$ potential. Like the Calogero model 
it has a linear spectrum and the eigenfunctions are:
\begin{align}
  \mathcal{E}_n&=n/2, \quad n=0,1,2,\ldots,\\
  \phi_0(x)&=\sqrt{\frac{\prod_{j=1}^3\Gamma(a_j+ix)}{\Gamma(2ix)}
  \frac{\prod_{j=1}^3\Gamma(a_j-ix)}{\Gamma(-2ix)}}\,,
  \quad \eta(x)=x^2,
  \label{phi0cdH}\\
  \phi_n(x)&=\phi_0(x)P_n(x^2),\quad
  P_n(\eta)\eqdef S_n(\eta\,;a_1,a_2,a_3),
  \label{phincdH}
\end{align}
in which $S_n(\eta\,;a_1,a_2,a_3)$ is the continuous dual Hahn
polynomial \eqref{defcdH}.
For deriving the classical solution, let us note that the quantum
potential \eqref{dualconthahnpot} has acquired quantum corrections from
the classical one:
\begin{equation}
  V_c(x)=\frac{\prod_{j=1}^3(a_j+ix)}{(2ix)^2}\,.
  \label{cldualconthahnpot}
\end{equation}
The classical motion is simple:
\begin{align}
  \{\mathcal{H}_c,x^2\}&=-\frac{\sqrt{\prod_{j=1}^3(a_j^2+x^2)}}{2x}
  \sinh p,\\
  \{\mathcal{H}_c,\{\mathcal{H}_c,x^2\}\}
  &=-x^2/4+ 2\mathcal{H}_c^2+b_1\mathcal{H}_c+b_2/4,\quad
  b_1\eqdef\!\sum_{1\leq j\leq 3}a_j\,,\quad 
  b_2\eqdef\!\!\sum_{1\leq j<k\leq 3}a_ja_k\,,\\
  x^2(t)&=\bigl(x^2(0)-8\mathcal{H}_{c0}^2-4b_1\mathcal{H}_{c0}-b_2\bigr)
  \cos[t/2]+8\mathcal{H}_{c0}^2+4b_1\mathcal{H}_{c0}+b_2 \n
  &\quad +\frac{\sqrt{\prod_{j=1}^3(a_j^2+x^2(0))}}{x(0)}\,\sinh p(0)\,
  \sin[t/2].
\end{align}

The quantum version is almost the same:
\begin{align}
  [\mathcal{H},x^2]&=-ix(T_+-T_-)-(T_++T_-)/2,\\
  [\mathcal{H},[\mathcal{H},x^2]\,]&=
  x^2/4 +R_{-1}(\mathcal{H}),\quad 
  R_{-1}(\mathcal{H})
  =-\bigl(2\mathcal{H}^2+(b_1-1/2)\mathcal{H}+b_2/4\bigr),\\ 
  e^{it\mathcal{H}}\,x^2\,e^{-it\mathcal{H}}&=
  2i[\mathcal{H},x^2]\sin[t/2]
  +\bigl(x^2 +4R_{-1}(\mathcal{H})\bigr)\cos[t/2]-4R_{-1}(\mathcal{H}).
\end{align}
The annihilation and creation operators are:
\begin{align}
  a^{(\pm)}&=\pm[\mathcal{H},x^2]+x^2/2+2R_{-1}(\mathcal{H})\n
  &=\mp ix(T_+-T_-)\mp (T_++T_-)/2+x^2/2+2R_{-1}(\mathcal{H}).
\end{align}
When applied to the eigenvector $\phi_n$, we obtain:
\begin{align}
  a^{(-)}\phi_n&=-n\prod_{1\leq j<k\leq 3}(n+a_j+a_k-1)\cdot\phi_{n-1},\\
  a^{(+)}\phi_n&=-\phi_{n+1}.
\end{align}
The similarity transformed operators are:
\begin{align}
  \phi_0(x)^{-1}\ a^{(\pm)}\ \phi_0(x)&=
  x^2/2-4\tilde{\mathcal{H}}^2-2(b_1-1/2)\tilde{\mathcal{H}}-b_2/2\n
  &\quad \mp\bigl((1/2+ix)V(x)e^p+(1/2-ix)V(x)^*e^{-p}\bigr),
\end{align}
in which $\tilde{\mathcal{H}}=\phi_0(x)^{-1}\mathcal{H}\phi_0(x)
=(V(x)e^p+V(x)^*e^{-p}-V(x)-V(x)^*)/2$ is the Hamiltonian counterpart
at the polynomial level satisfying 
$\tilde{\mathcal{H}}P_n(x^2)=n/2\,P_n(x^2)$, see Appendix C.
The coherent state is
\begin{equation}
  \psi(x)=\phi_0(x)\sum_{n=0}^\infty
  \frac{(-\lambda)^n}{n!\,\prod_{1\leq j<k\leq 3}(a_j+a_k)_n}\,P_n(x^2).
\end{equation}
We do not know if this sum has a concise expression or not.
The commutation relations among $\mathcal{H}$, and $a^{(\pm)}$
are more complicated than $\mathfrak{su}(1,1)$:
\begin{align}
  [\mathcal{H},a^{(\pm)}]&=\pm a^{(\pm)}/2,\n
  {}[a^{(-)},a^{(+)}]&=32\mathcal{H}^3+24(b_1-1/2)\mathcal{H}^2
  +\bigl(4(b_1-1/2)^2+4b_2+1\bigr)\mathcal{H}+b_1b_2-a_1a_2a_3.
\end{align}

As in the Meixner-Pollaczek case \eqref{acfact}, the annihilation 
and creation operators for the continuous dual Hahn polynomial
factorise into the operators $\mathcal{A}$ and
$\mathcal{A}^\dagger$ appearing in the Hamiltonian
$\mathcal{H}=\mathcal{A}^\dagger\mathcal{A}/2$: 
\begin{equation}
  a^{(-)}=X^\dagger\mathcal{A},\quad
  a^{(+)}=\mathcal{A}^\dagger X. 
  \label{cdHacfac}
\end{equation}
The operator $X$ in this case reads
\begin{align}
  X&=-iS_+T_+ +\Bigl(x-iV(x-\tfrac{i}{2})^*
  -i\,\frac{\prod_{j=1}^3(2a_j-1)}{8(1+x^2)}\Bigr)S_+ \n
  &\quad +iS_-T_- +\Bigl(x+iV(x-\tfrac{i}{2})\,\,
  +i\,\frac{\prod_{j=1}^3(2a_j-1)}{8(1+x^2)}\Bigr)S_-.
  \label{XcdH}
\end{align}
These $X$ and $X^{\dagger}$ compensate the shift of the parameters
$(a_1,a_2,a_3)$ caused by $\mathcal{A}^\dagger$ and $\mathcal{A}$, 
respectively.
See Appendix {\bf B} for more details.

\subsubsection{Wilson polynomial}
\label{wilson}

The Wilson polynomial has four parameters ($a_1,a_2,a_3,a_4$)
and is considered as a three parameter deformation of the Laguerre 
polynomial $L_n^{(\alpha)}$. 
The factorised Hamiltonian \eqref{H} of the Wilson polynomial
has a potential function $V$:
\begin{equation}
  V(x)=\frac{\prod_{j=1}^4(a_j+ix)}{2ix(2ix+1)},\quad
  0<x<\infty,\quad a_1,a_2,a_3,a_4>0.
  \label{wilsonpot}
\end{equation}
As a dynamical system this is another deformation of the Calogero model,
or a deformed $x^2+1/x^2$ potential.
The spectrum is now quadratic in $n$ and the eigenfunctions are:
\begin{align}
  \mathcal{E}_n&=n\bigl(n+\mbox{$\sum_{j=1}^4a_j$}-1\bigr)/2,
  \quad n=0,1,2,\ldots,\\
  \phi_0(x)&=\sqrt{\frac{\prod_{j=1}^4\Gamma(a_j+ix)}{\Gamma(2ix)}
  \frac{\prod_{j=1}^4\Gamma(a_j-ix)}{\Gamma(-2ix)}}\,,
  \quad \eta(x)=x^2,
  \label{phi0W}\\
  \phi_n(x)&=\phi_0(x)P_n(x^2),\quad
  P_n(\eta)\eqdef W_n(\eta\,;a_1,a_2,a_3,a_4),
  \label{phinW}
\end{align}
in which $W_n(\eta\,;a_1,a_2,a_3,a_4)$ is the Wilson polynomial \eqref{defW}.
The classical motion looks like a cross between those of the 
continuous Hahn and the continuous dual Hahn potentials with
the classical potential $V_c$:
\begin{align}
  \{\mathcal{H}_c,x^2\}&=-\frac{\sqrt{\prod_{j=1}^4(a_j^2+x^2)}}{2x}
  \sinh p,\quad V_c(x)=\frac{\prod_{j=1}^4(a_j+ix)}{(2ix)^2}\,
  \label{clwilsonpot}\\
  \{\mathcal{H}_c,\{\mathcal{H}_c,x^2\}\}
  &=-2x^2(\mathcal{H}_c+c_1)-R_{-1}(\mathcal{H}_c),\\
  R_{-1}(\mathcal{H}_c)&=-2(\mathcal{H}_c^2+c_2\mathcal{H}_c+c_3),\qquad
  c_1=b_1^2/8,\quad c_2=b_2,\quad c_3=b_1b_3/4,\\
  x^2(t)&=\Bigl(x^2(0)
  +\frac{R_{-1}(\mathcal{H}_{c0})}{2(\mathcal{H}_{c0}+c_1)}\Bigr)
  \cos\bigl[t\sqrt{2(\mathcal{H}_{c0}+c_1)}\,\bigr]
  -\frac{R_{-1}(\mathcal{H}_{c0})}{2(\mathcal{H}_{c0}+c_1)}\n
  &\quad +\frac{\sqrt{\prod_{j=1}^4(a_j^2+x^2(0))}}{2x(0)}\,\sinh p(0)\,
  \frac{\sin\bigl[t\sqrt{2(\mathcal{H}_{c0}+c_1)}\,\bigr]}
  {\sqrt{2(\mathcal{H}_{c0}+c_1)}} \,,
\end{align}
where we use the abbreviation
\begin{equation}
  b_1\eqdef\!\sum_{1\leq j\leq 4}a_j\,,\quad
  b_2\eqdef\!\!\sum_{1\leq j< k\leq 4}a_ja_k\,,\quad
  b_3\eqdef\!\!\!\sum_{1\leq j< k<l\leq 4}a_ja_ka_l\,.
\end{equation}

The quantum version has almost the same form with quantum corrections
in the coefficients $c_1,c_2$ and $c_3$:
\begin{align}
  [\mathcal{H},x^2]&=-ix(T_+-T_-)-(T_++T_-)/2,\\
  [\mathcal{H},[\mathcal{H},x^2]\,]&=
  [\mathcal{H},x^2] +2x^2(\mathcal{H}+c_1)+R_{-1}(\mathcal{H}),\\
  R_{-1}(\mathcal{H})&=-2(\mathcal{H}^2+c_2\mathcal{H}+c_3),\\
  &\qquad
  c_1=b_1(b_1-2)/8,\quad c_2=b_2-b_1/2,\quad c_3=(b_1-2)b_3/4,\\
  e^{it\mathcal{H}}\,x^2\,e^{-it\mathcal{H}}&=
  [\mathcal{H},x^2]
  \frac{e^{i\alpha_+(\mathcal{H})t}-e^{i\alpha_-(\mathcal{H})t}}
  {2\sqrt{2\mathcal{H}'}}
  -\frac{R_{-1}(\mathcal{H})}{2(\mathcal{H}+c_1)}\n
  &\quad+\Bigl(x^2+\frac{R_{-1}(\mathcal{H})}{2(\mathcal{H}+c_1)}\Bigr)
  \frac{-\alpha_-(\mathcal{H})e^{i\alpha_+(\mathcal{H})t}
  +\alpha_+(\mathcal{H})e^{i\alpha_-(\mathcal{H})t}}
  {2\sqrt{2\mathcal{H}'}}.\\
  \alpha_\pm(\mathcal{H})&=1/2\pm\sqrt{2\mathcal{H}'},\quad
  2\mathcal{H}'\eqdef 2\mathcal{H}+2c_1+1/4.
\end{align}
The annihilation and creation operators are:
\begin{align}
  a^{\prime(\pm)}=a^{(\pm)}2\sqrt{2\mathcal{H}'}
  &=\pm [\mathcal{H},x^2]\mp x^2\alpha_{\mp}(\mathcal{H})
  +\frac{R_{-1}(\mathcal{H})}{\sqrt{2\mathcal{H}'}\pm1/2}\n
  &=\mp ix(T_+-T_-)\mp(T_++T_-)/2\mp x^2\alpha_{\mp}(\mathcal{H})
  +\frac{R_{-1}(\mathcal{H})}{\sqrt{2\mathcal{H}'}\pm1/2}\,.
\end{align}
When applied to the eigenvector $\phi_n$, we obtain as
$2\mathcal{E}_n+2c_1+1/4=(2n+b_1-1)^2/4$:
\begin{align}
  a^{\prime(-)}\phi_n&=
  -\frac{n\prod_{1\leq j<k\leq 4}(n+a_j+a_k-1)}{(2n+b_1-2)(2n+b_1-1)}\,
  \phi_{n-1}\,,\\
  a^{\prime(+)}\phi_n&=-\frac{n+b_1-1}{(2n+b_1-1)(2n+b_1)}\,\phi_{n+1}\,.
\end{align}
The similarity transformed operators act as
\begin{align}
  &\phi_0(x)^{-1}a^{\prime(\pm)}\phi_0(x)\cdot P_n(x^2)\\
  =&\ \Bigl(\pm(\mathcal{E}_n-\mathcal{E}_{n\mp 1})x^2
  \pm\frac{R_{-1}(\mathcal{E}_n)}{\mathcal{E}_{n\pm 1}-\mathcal{E}_n}
  \mp\bigl((1/2+ix)V(x)e^p+(1/2-ix)V(x)^*e^{-p}\bigr)\Bigr)
  P_n(x^2).\nonumber
\end{align}
The coherent state $\psi$ and $\psi'$ are
\begin{align}
  \psi(x)&=\phi_0(x)\sum_{n=0}^\infty\frac{(-\lambda)^n}{n!}\,
  \frac{(a_1+a_2+a_3+a_4)_{2n}}{\prod_{1\leq j<k\leq 4}(a_j+a_k)_n}\, 
  P_n(x^2),\\
  \psi'(x)&=\phi_0(x)\sum_{n=0}^\infty\frac{(-2\lambda)^n}{n!}\,
  \frac{((a_1+a_2+a_3+a_4)/2)_n}{\prod_{1\leq j<k\leq 4}(a_j+a_k)_n}\, 
  P_n(x^2).
\end{align}

\section{Summary and Comments}
\label{comments}
\setcounter{equation}{0}

Unified theory of annihilation-creation operators $a^{(\pm)}$
is developed for various exactly solvable quantum mechanical 
systems possessing the `sinusoidal coordinate'.
It applies to most of the degree one solvable quantum mechanical systems
as well as the solvable `discrete' quantum mechanical systems,
which are also shape-invariant \cite{os4}.
The eigenfunctions of the latter are described by the Askey-scheme of
hypergeometric orthogonal polynomials \cite{koeswart}.
The method provides an independent {\em algebraic solution\/}
of these quantum systems. The energy spectrum is obtained \`a la 
Heisenberg and Pauli from the Heisenberg operator solution 
for the `sinusoidal coordinate'
$\eta$, $e^{it\mathcal{H}}\eta e^{-it\mathcal{H}}$ and 
the entire eigenfunctions are explicitly obtained as
$\{(a^{(+)})^n\phi_0\}$, $n=0,1,\ldots$, in which $\phi_0$ is
determined by $a^{(-)}\phi_0=0$.
Various examples are worked out in section \ref{gentheory} and \ref{3rd}.
It also applies to theories with a finite number of bound states.
It should be stressed that these annihilation-creation operators are
{\em natural\/} ones containing the differential (difference)
operators, in contradistinction to those annihilation-creation operators
introduced in the algebraic theory of coherent states \cite{coherents}.
By a similarity transformation in terms of the ground state wavefunction
$\phi_0$, the Heisenberg operator solution gives the structure relation
for the corresponding orthogonal polynomials \cite{koorn} and 
the annihilation-creation operators provide their shift down-up operators.
Another characteristic feature is the uniqueness.
Except for the overall factor, which is intrinsically undetermined,
the action $a^{(\pm)}\phi_n$ is completely determined by the Hamiltonian
of the system.
This means that the relative weights of the terms in $a^{(\pm)}\phi_n$
are governed by the energy spectrum. 
We have shown in some detail that this type of algebraic exact
solvability is valid at both classical and quantum levels.
This is in good contrast with shape-invariance, which is a strictly 
quantum notion.
The necessary and sufficient condition for the existence of the
`sinusoidal coordinate' is worked out for the ordinary quantum mechanical
systems in Appendix {\bf A}. It is a good challenge to derive a
corresponding result for the `discrete' quantum mechanical systems.

\bigskip

Generalisation of the present formalism to multi-particle
systems is highly desirable.
Simplest multi-particle systems possessing the `sinusoidal coordinate'
and the corresponding Heisenberg operator solution is the 
Calogero systems based on any root system \cite{kps}.
In fact a more general Hamiltonian 
\begin{align}
  &\mathcal{H}=\frac{1}{2}\sum_{j=1}^n(p_j^2+x_j^2)+V(x),\quad
  \sum_{j=1}^n x_j\frac{\partial}{\partial x_j}V(x)=-2V(x),\\
  &[\mathcal{H},[\mathcal{H},\eta]]=4(\eta-\mathcal{H}),
  \quad \eta=\sum_{j=1}^nx_j^2,
\end{align}
of harmonic oscillators modified by 
a generic homogeneous degree -2 potential has the same property.
The corresponding eigenfunctions are the Laguerre polynomials
again \cite{gamba, kps}.
As is well known the annihilation-creation operators of the harmonic
oscillator have a quite wide applicability in many branches of physics.
We wonder if the newly found annihilation-creation operators for the other
solvable quantum mechanical  systems might find an equally wide range
of applications.

\section*{Acknowledgements}

We thank F.\, Calogero for stimulating discussion.
This work is supported in part by Grant-in-Aid for Scientific
Research from the Ministry of Education, Culture, Sports, Science and
Technology, No.18340061 and No.16340040.

\renewcommand{\thesection}{\Alph{section}}
\setcounter{section}{1}
\renewcommand{\theequation}{A.\arabic{equation}}
\setcounter{equation}{0}
\section*{Appendix A: Determination of the potentials having
the `sinusoidal coordinate'}

We have seen that the existence of the `sinusoidal coordinate'
or the exact Heisenberg operator solution \eqref{quantsol} leads to
the unified definition of the annihilation-creation operators.
All the examples discussed in the text share the common property
of `shape-invariance', thanks to which the corresponding quantum 
systems are exactly solvable.
Here in Appendix {\bf A} we analyse, within the context of ordinary
quantum mechanics, the necessary and sufficient condition for the
existence of the `sinusoidal coordinate' and show that such systems
constitute a sub-group of known `shape-invariant' quantum mechanics. 
For the `discrete' quantum mechanical systems, writing down
corresponding conditions is easy. It would be a good challenge to
provide a complete list of `discrete' quantum mechanical systems
admitting the `sinusoidal coordinate' or the exact Heisenberg 
operator solution.

For a given pair $(\eta(x), \mathcal{H})$ of a coordinate function
$\eta(x)$ and a Hamiltonian $\mathcal{H}$ to satisfy the exact 
Heisenberg operator solution \eqref{quantsol} is equivalent to 
the condition \eqref{twocom} that the multiple commutators of 
$\mathcal{H}$ with $\eta$ form a closed algebra at level two
\begin{equation}
  ({\rm ad}\,\mathcal{H})^2\eta(x)\equiv
  [\mathcal{H},[\mathcal{H},\eta(x)]\,]=\eta\,
  R_0(\mathcal{H})+[\mathcal{H},\eta(x)]
  \,R_1(\mathcal{H})+R_{-1}(\mathcal{H}).
  \label{quadclose}
\end{equation}
Here the coefficients $R_0(\mathcal{H})$, $R_1(\mathcal{H})$
and $R_{-1}(\mathcal{H})$ are polynomials in the Hamiltonian
$\mathcal{H}$.
It should be stressed that this condition is purely algebraic and
the knowledge that the eigenfunctions have the general structure
\eqref{special} is irrelevant. The latter \eqref{special} is
a consequence of the condition \eqref{quadclose}.
For the ordinary quantum mechanical system with potential $V(x)$
\begin{equation}
  \mathcal{H}=-\frac12\frac{d^2}{dx^2}+V(x),
\end{equation}
the commutator between $\mathcal{H}$ and $\eta$ reads
\begin{align}
  [\mathcal{H},\eta]&=-\eta'\frac{d}{dx}-\frac12\eta'',
  \label{hetacom}\\
  \bigl({\rm ad}\mathcal{H}\bigr)^2\eta&=\eta''\frac{d^2}{dx^2}
  +\eta'''\frac{d}{dx}+\frac14\eta''''+\eta' V',
  \label{adH^2eta}
\end{align}
in which primes denote differentiation with respect to $x$.
{}From \eqref{adH^2eta} we see that the l.h.s. of
\eqref{quadclose} contains the derivative operator (the momentum
operator) at most quadratic degree.
So must be the r.h.s. since the momentum operator can come in as a
part of $\mathcal{H}$ (as $p^2/2$) or as $ [\mathcal{H},\eta]$, 
see \eqref{hetacom}.
Then we can parametrise
\begin{equation}
  R_0(\mathcal{H})=r_0^{(1)}\mathcal{H}+r_0^{(0)},\quad
  R_1(\mathcal{H})=r_1,\quad
  R_{-1}(\mathcal{H})=r_{-1}^{(1)}\mathcal{H}+r_{-1}^{(0)},\
\end{equation}
in which $r_j^{(k)}$ are all constants. Then the coefficients of
the operators $\frac{d^2}{dx^2}$, $\frac{d}{dx}$ and the function part of
\eqref{quadclose} give the conditions:
\begin{align}
  \frac{d^2\eta}{dx^2}&=-\frac12(r_0^{(1)}\eta+r^{(1)}_{-1})\,,
  \label{acond1}\\
  \frac{d^3\eta}{dx^3}&=-r_1\frac{d\eta}{dx}\,,
  \label{acond2}\\
  \frac14\frac{d^4\eta}{dx^4}+\frac{d\eta}{dx}\frac{dV}{dx}
  &=-\frac12r_1\frac{d^2\eta}{dx^2}+(r_0^{(1)}\eta+r^{(1)}_{-1})V
  +r^{(0)}_0\eta+r^{(0)}_{-1}\,.
  \label{acond3}
\end{align}
The first condition \eqref{acond1} simply means that $\eta(x)$ is
either a {\em trigonometric\/} or a {\em hyperbolic\/} function of
$x$ which gives an {\em exponential\/} function or a 
{\em quadratic\/} and {\em linear  polynomial\/} in $x$ in the
degenerate limits. 
By comparing \eqref{acond1} and \eqref{acond2} we obtain
\begin{equation}
  r_0^{(1)}=2r_1.
\end{equation}
Then \eqref{acond3} reduces to
\[
   \frac{d\eta}{dx}\frac{dV}{dx}
  +\frac{d^2\eta}{dx^2}\Bigl(2V+\frac14r_1\Bigr)
  =r^{(0)}_0\eta+r^{(0)}_{-1},
\]
which integrates easily when multiplied by $d\eta/dx$:
\begin{equation}
  \Bigl(\frac{d\eta}{dx}\Bigr)^2\Bigl(V+\frac{r_1}{8}\Bigr)
  =\frac{r^{(0)}_0}{2}\eta^2+r^{(0)}_{-1}\eta+c.
  \label{vint}
\end{equation}
Here $c$ is the constant of integration.
Thus we have determined the possible form of the potential $V$
in terms of the `sinusoidal coordinate' $\eta(x)$ and its first
derivative $d\eta/dx$, with five parameters $r_1$,
$r_0^{(0)}$, $r_{-1}^{(1)}$, $r_{-1}^{(0)}$ and $c$ in \eqref{vint}
and two more possible constants of integration of \eqref{acond1}:
\begin{equation}
  V(x)=\frac{1}{(\frac{d\eta}{dx})^2}
  \Bigl(\frac{r^{(0)}_0}{2}\eta^2+r^{(0)}_{-1}\eta+c\Bigr)-\frac{r_1}{8}.
  \label{cond3p}
\end{equation}
The actual number of essentially free parameters is much less, 
since the origin of the quadratic potential, or the location of 
the singularity, etc, could be freely adjusted by introducing new
variable $x_{\rm new}=\alpha x+\beta$ 
($\mathcal{H}_{\rm new}=\mathcal{H}/\alpha^2$).
The condition that the Hamiltonian must be bounded from below imposes
some constraints on the parameters.
The overall additive constant is fixed uniquely
when the ground state energy is required to be vanishing 
$\mathcal{E}_0=0$.

It is rather straightforward to determine all the potentials
possessing the `sinusoidal coordinate' and thus algebraically
exactly solvable. They all belong to the known group of
shape-invariant potentials. 
Except for the trivial case $V=0$, we have
\begin{enumerate}
\item
Rational case, $r_1=0$. The generic solution of \eqref{acond1} is
\begin{equation}
  \eta(x)=-\frac{1}{4}r_{-1}^{(1)}x^2+c_1x+c_2,
\end{equation}
with $c_1$ and $c_2$ being the constant of integration.
Two special cases are of interest:
$\eta(x)=x$ gives the harmonic oscillator and $\eta(x)=x^2$ leads 
to the $x^2+1/x^2$ potential discussed in section \ref{calo}.
\item
Trigonometric case, $r_1>0$.
The generic solution of \eqref{acond1} is
\begin{eqnarray}
  \eta(x)=-\frac{r_{-1}^{(1)}}{2r_1}+c_1\cos\sqrt{r_1}\,x
  +c_2\sin\sqrt{r_1}\,x,
\end{eqnarray}
with $c_1$ and $c_2$ being real constants of integration
due to the reality (hermiticity) of $\eta$.
By rescaling and shift of the coordinate $x$, it reduces to 
the P\"oschl-Teller potential discussed in section \ref{poshtell}. 
The $1/\sin^2x$ potential in section \ref{sutpot} and the symmetric top
 are obtained as degenerate cases.
\item
Hyperbolic and exponential cases, $r_1<0$.
The generic solution of \eqref{acond1} is
\begin{eqnarray}
  \eta(x)=-\frac{r_{-1}^{(1)}}{2r_1}+c_1\cosh\sqrt{-r_1}\,x
  +c_2\sinh\sqrt{-r_1}\,x,
\end{eqnarray}
in which the constants of integration $c_1$ and $c_2$ could be
vanishing or equal $c_1=\pm c_2$.
The generic case leads to the hyperbolic P\"oschl-Teller potential
and the degenerate cases contain the soliton potential in section
\ref{soliton} and hyperbolic symmetric tops and the Morse potential
in section \ref{morse}, etc. 
We could not discuss all due to space limitation.
\end{enumerate}

In all these examples, the prepotential $W$ has also a simple expression
in terms of $\eta$ and $d\eta/dx$:
\begin{equation}
  \frac{dW}{dx}=\frac{a\eta+b}{\frac{d\eta}{dx}},
  \qquad a=-\sqrt{r_0^{(0)}+r_1^2/4},
  \quad b=\frac{2r_{-1}^{(0)}}{2a+r_1}+\frac{r_{-1}^{(1)}}{4}.
\end{equation}
Here the prepotential $W$ is related to the ground state wave function
$\phi_0$ and thus to the potential $V$ as
\[
  \phi_0(x)=e^{W(x)},\quad
  V=\frac{1}{2}\biggl(\Bigl(\frac{dW}{dx}\Bigr)^2+\frac{d^2W}{dx^2}\biggr),
\]
and it plays an important role in supersymmetric (shape-invariant) quantum
mechanics \cite{susyqm,bms}.

It should be stressed that not all shape-invariant and exactly
solvable potentials admit the `sinusoidal coordinate'. 
Such examples are the Kepler problems in rational, 
spherical and hyperbolic coordinates and the Rosen-Morse potential,
respectively:
\begin{align}
  V(x)&=\frac12\Bigl(-\frac{2}{x}+\frac{g(g-1)}{x^2}
  +\frac{1}{g^2}\Bigr)\,,\\
  V(x)&=\frac12\Bigl(-2\mu\cot x+\frac{g(g-1)}{\sin^2x}
  +\frac{\mu^{\,2}}{g^2}-g^2\Bigr)\,,\\
  V(x)&=\frac12\Bigl(-2\mu\coth x+\frac{g(g-1)}{\sinh^2x}
  +\frac{\mu^{\,2}}{g^2}+g^2\Bigr)\,,\\
  V(x)&=\frac12\Bigl(2\mu\tanh x-\frac{g(g+1)}{\cosh^2x}
  +\frac{\mu^{\,2}}{g^{2}}+g^{2}\Bigr).
\end{align}
Their wavefunctions do not have the general structure \eqref{special}, 
either.

\section*{Appendix B: Interpretation in terms of Shape Invariance}
\label{simtran}
\renewcommand{\theequation}{B.\arabic{equation}}
\setcounter{equation}{0}

As shown in section \ref{gentheory}, the annihilation-creation operators
are completely determined once the closed relationship \eqref{twocom} 
among $\eta$, $[\mathcal{H},\eta]$ and 
$[\mathcal{H},[\mathcal{H},\eta]]$ is obtained.
Although it plays no active role in the determination of the
annihilation-creation operators, {\em shape invariance\/} is the common
property underlying all these exactly solvable Hamiltonians discussed 
in this paper. 
Therefore it is interesting as well as illuminating to understand 
the mechanism of the annihilation-creation operators within the 
framework of shape invariance.
For this purpose we concentrate on the annihilation-creation operators 
of the Meixner-Pollaczek polynomials \eqref{acfact} and of the 
continuous dual Hahn polynomials \eqref{cdHacfac}, which factorise 
into the operators $\mathcal{A}$ and $\mathcal{A}^\dagger$ constituting 
the shape invariant Hamiltonian
$\mathcal{H}=\mathcal{A}^\dagger\mathcal{A}/2$.
Another motivation of this Appendix is to provide a bridge between 
the physics of `discrete' quantum mechanics \cite{os4} and the 
analysis of Askey-scheme of hypergeometric polynomials \cite{koeswart}. 
The latter focuses on the polynomial part of the eigenfunctions, 
whose orthogonal measure is provided by the ground state 
wavefunction \eqref{measure}.

Let us start with recapitulating the rudimentary facts of the 
shape-invariant `discrete' quantum mechanics as developed in \cite{os4}.
Knowledgeable readers may jump to the main results
\eqref{mpxsim}--\eqref{cdhxdagsim}, but some intermediate results 
\eqref{groundprop} and \eqref{pSp}--\eqref{pTp} would  also be interesting
in connection with the `sinusoidal coordinate' $\eta(x)$.

A shape invariant quantum mechanical system consists of a series of 
isospectral Hamiltonians $\{\mathcal{H}(\bm{\lambda})\}$ parametrised 
by (a set of ) parameters $\bm{\lambda}=(\lambda_1,\lambda_2,\cdots)$: 
$$
  \mathcal{H}(\bm{\lambda})=\mathcal{A}(\bm{\lambda})^{\dagger}
  \mathcal{A}(\bm{\lambda})/2,\quad
  \phi_n(x\,;\bm{\lambda})
  =\phi_0(x\,;\bm{\lambda})P_n(\eta(x)\,;\bm{\lambda}),\quad
  \mathcal{E}_n(\bm{\lambda}),\quad\mbox{etc.}
$$
Shape invariance is tersely expressed as
\begin{equation}
  \mathcal{A}(\bm{\lambda})\mathcal{A}(\bm{\lambda})^{\dagger}
  =\mathcal{A}(\bm{\lambda}+\bm{\delta})^{\dagger}
  \mathcal{A}(\bm{\lambda}+\bm{\delta})+2\mathcal{E}_1(\bm{\lambda}),
\end{equation}
where $\bm{\delta}$ is a shift of the parameter.\footnote{
In the case of the Askey-Wilson polynomials, this is modified to
$\mathcal{A}(\bm{\lambda})\mathcal{A}(\bm{\lambda})^{\dagger}
 =q^{2\delta'}\mathcal{A}(\bm{\lambda}*q^{\bm{\delta}})^{\dagger}
 \mathcal{A}(\bm{\lambda}*q^{\bm{\delta}})+2\mathcal{E}_1(\bm{\lambda})$,
where $\delta'$ is a constant and $\bm{\lambda}*q^{\bm{\delta}}=
(\lambda_1q^{\delta_1},\lambda_2q^{\delta_2},\cdots)$.
}
The operator $\mathcal{A}(\bm{\lambda})$ maps the eigenvectors of 
$\mathcal{H}(\bm{\lambda})$ to those of 
$\mathcal{H}(\bm{\lambda}+\bm{\delta})$ and the other operator 
$\mathcal{A}(\bm{\lambda})^\dagger$ acts in the opposite direction.
In each case studied in this paper, the parameter $\bm{\lambda}$ and 
the shift $\bm{\delta}$ are
\begin{alignat}{2}
  &\mbox{Meixner-Pollaczek} &\ :\ \,
  \bm{\lambda}&=a,\quad \bm{\delta}=1/2,\\
  &\mbox{continuous Hahn} &\ :\ \,
  \bm{\lambda}&=(a_1,a_2),\quad \bm{\delta}=(1/2,1/2),\\
  &\mbox{continuous dual Hahn} &\ :\ \,
  \bm{\lambda}&=(a_1,a_2,a_3),\quad \bm{\delta}=(1/2,1/2,1/2),\\
  &\mbox{Wilson} &\ :\ \,
  \bm{\lambda}&=(a_1,a_2,a_3,a_4),\quad \bm{\delta}=(1/2,1/2,1/2,1/2),\\
  &\mbox{Askey-Wilson} &\ :\ \,
  \bm{\lambda}&=(a_1,a_2,a_3,a_4),\quad \bm{\delta}=(1/2,1/2,1/2,1/2),
  \quad \delta'=-1/2.
\end{alignat}
The ground state $\phi_0(x)$ and the orthogonal polynomial 
$P_n(\eta(x))$ are given in \eqref{phi0MP}--\eqref{phinMP},
\eqref{phi0cH}--\eqref{phincH}, \eqref{phi0cdH}--\eqref{phincdH},
\eqref{phi0W}--\eqref{phinW} and \eqref{phi0AW}--\eqref{phinAW}.
These ground states satisfy\footnote{
For the Askey-Wilson polynomials, this relation reads
$\phi_0(x-i\gamma/2\,;\bm{\lambda}*q^{\bm{\delta}})
=\sqrt{V(z\,;\bm{\lambda})}\,\varphi(x-i\gamma/2)\phi_0(x\,;\bm{\lambda})$,
where $\gamma=\log q$.
}
\begin{equation}
  \phi_0(x-i/2\,;\bm{\lambda}+\bm{\delta})
  =\sqrt{V(x\,;\bm{\lambda})}\,\varphi(x-i/2)\phi_0(x\,;\bm{\lambda}),
  \label{groundprop}
\end{equation}
where $\varphi(x)\propto \eta'(x)$ is given by
\begin{alignat}{2}
  &\mbox{Meixner-Pollaczek} &\ :\ \,
  \varphi(x)&=1,\\
  &\mbox{continuous Hahn} &\ :\ \,
  \varphi(x)&=1,\\
  &\mbox{continuous dual Hahn} &\ :\ \,
  \varphi(x)&=2x,\\
  &\mbox{Wilson} &\ :\ \,
  \varphi(x)&=2x,\\
  &\mbox{Askey-Wilson} &\ :\ \,
  \varphi(x)&=-2\sin x=i(z-z^{-1}).
\end{alignat}

Let us consider $S_{\pm}(\bm{\lambda})$, $T_{\pm}(\bm{\lambda})$, 
$\mathcal{A}(\bm{\lambda})$ given in \eqref{spmdef}--\eqref{aaddef}.
By using the property \eqref{groundprop}, we have
\begin{align}
  \phi_0(x\,;\bm{\lambda}+\bm{\delta})^{-1}\,S_{\pm}(\bm{\lambda})\,\,
  \phi_0(x\,;\bm{\lambda})&=\varphi(x)^{-1}\,e^{\pm p/2}\,,
  \label{pSp}\\
  \phi_0(x\,;\bm{\lambda})^{-1}\,S_{\pm}(\bm{\lambda})^{\dagger}\,\,
  \phi_0(x\,;\bm{\lambda}+\bm{\delta})&=
  \begin{cases}
   V(x\,;\bm{\lambda})\,e^{p/2}\,\varphi(x)\\
   V(x\,;\bm{\lambda})^*\,e^{-p/2}\,\varphi(x)\,.
  \end{cases}
  \label{pSdp}
\end{align}
(In the case of the Askey-Wilson polynomials, the following replacement is
needed:
$\bm{\lambda}+\bm{\delta}\Rightarrow\bm{\lambda}*q^{\bm{\delta}}$, 
$e^{\pm p/2}\Rightarrow q^{\pm D/2}$, 
$V(x\,;\bm{\lambda})\Rightarrow V(z\,;\bm{\lambda})$.
Hereafter we will omit similar remarks.)
{}From this, we obtain
\begin{align}
  F(\bm{\lambda})&\eqdef
  \phi_0(x\,;\bm{\lambda}+\bm{\delta})^{-1}\,\mathcal{A}(\bm{\lambda})\,\,
  \phi_0(x\,;\bm{\lambda})=i\,\varphi(x)^{-1}
  \bigl(e^{p/2}-e^{-p/2}\bigr)\,,\\
  B(\bm{\lambda})&\eqdef
  \phi_0(x\,;\bm{\lambda})^{-1}\,\mathcal{A}(\bm{\lambda})^{\dagger}\,\,
  \phi_0(x\,;\bm{\lambda}+\bm{\delta})=-i
  \bigl(V(x\,;\bm{\lambda})\,e^{p/2}-V(x\,;\bm{\lambda})^*\,e^{-p/2}\bigr)
  \varphi(x)\,,\\
  \widetilde{T}_{\pm}(\bm{\lambda})&\eqdef
  \phi_0(x\,;\bm{\lambda})^{-1}\,T_{\pm}(\bm{\lambda})\,\,
  \phi_0(x\,;\bm{\lambda})=
  \begin{cases}
   V(x\,;\bm{\lambda})\,e^{p}\\
   V(x\,;\bm{\lambda})^*\,e^{-p}\,.
  \end{cases}
  \label{pTp}
\end{align}
Therefore the similarity transformed Hamiltonian is
\begin{align}
  \widetilde{\mathcal{H}}(\bm{\lambda})&\eqdef
  \phi_0(x\,;\bm{\lambda})^{-1}\,\mathcal{H}(\bm{\lambda})\,\,
  \phi_0(x\,;\bm{\lambda})
  =B(\bm{\lambda})F(\bm{\lambda})/2\n
  &=\bigl(\widetilde{T}_+(\bm{\lambda})+\widetilde{T}_-(\bm{\lambda})
  -V(x\,;\bm{\lambda})-V(x\,;\bm{\lambda})^*\bigr)/2\,.
\end{align}
which acts on $P_n(\eta(x)\,;\bm{\lambda})$ as
$\tilde{H}(\bm{\lambda})P_n(\eta(x)\,;\bm{\lambda})
=\mathcal{E}_n(\bm{\lambda})P_n(\eta(x)\,;\bm{\lambda})$.

The forward shift operator $F(\bm{\lambda})$ and 
backward shift operator $B(\bm{\lambda})$ act on 
$P_n(\eta\,;\bm{\lambda})$ as
\begin{align}
  F(\bm{\lambda})P_n(\eta\,;\bm{\lambda})
  &=f_n(\bm{\lambda})P_{n-1}(\eta\,;\bm{\lambda}+\bm{\delta}),\\
  B(\bm{\lambda})P_n(\eta\,;\bm{\lambda}+\bm{\delta})
  &=b_n(\bm{\lambda})P_{n+1}(\eta\,;\bm{\lambda}),
\end{align}
where $f_n(\bm{\lambda})$ and $b_n(\bm{\lambda})$ are constants
 satisfying the relation  $f_n(\bm{\lambda})b_{n-1}(\bm{\lambda})/2
=\mathcal{E}_n(\bm{\lambda})$:
\begin{alignat}{2}
  &\mbox{Meixner-Pollaczek} &\ :\ \,
  f_n(\bm{\lambda})&=2,\quad b_n(\bm{\lambda})=n+1,\\
  &\mbox{continuous Hahn} &\ :\ \,
  f_n(\bm{\lambda})&=n+2a_1+2a_2-1,\quad b_n(\bm{\lambda})=n+1,\\
  &\mbox{continuous dual Hahn} &\ :\ \,
  f_n(\bm{\lambda})&=-n,\quad b_n(\bm{\lambda})=-1,\\
  &\mbox{Wilson} &\ :\ \,
  f_n(\bm{\lambda})&=-n(n+\mbox{$\sum_{j=1}^4a_j$}-1),\quad 
  b_n(\bm{\lambda})=-1,\\
  &\mbox{Askey-Wilson} &\ :\ \,
  f_n(\bm{\lambda})&=-q^{n/2}(q^{-n}-1)(1-a_1a_2a_3a_4q^{n-1}),
  \quad b_n(\bm{\lambda})=-q^{-(n+1)/2}.
\end{alignat}

For the Meixner-Pollaczek and the continuous dual Hahn cases, 
 we have seen that the annihilation-creation operators are
factorised $a^{(-)}\propto X^\dagger\mathcal{A}$ and 
$a^{(+)}\propto \mathcal{A}^\dagger X$, \eqref{acfact}, \eqref{cdHacfac}.
By using \eqref{pSp}--\eqref{pSdp} and \eqref{pTp}, 
$\phi_0(x\,;\bm{\lambda}+\bm{\delta})^{-1}X(\bm{\lambda})
\phi_0(x\,;\bm{\lambda})$ and
$\phi_0(x\,;\bm{\lambda})^{-1}$ $X(\bm{\lambda})^{\dagger}
\phi_0(x\,;\bm{\lambda}+\bm{\delta})$ can be written down explicitly.
They act on $P_n(\eta\,;\bm{\lambda})$ as
for the Meixner-Pollaczek case:
\begin{align}
  \phi_0(x\,;\bm{\lambda}+\bm{\delta})^{-1}\,X(\bm{\lambda})\,\,
  \phi_0(x\,;\bm{\lambda})\cdot P_n(\eta\,;\bm{\lambda})
  &=2P_n(\eta\,;\bm{\lambda}+\bm{\delta}),
  \label{mpxsim}\\
  \phi_0(x\,;\bm{\lambda})^{-1}\,X(\bm{\lambda})^{\dagger}\,\,
  \phi_0(x\,;\bm{\lambda}+\bm{\delta})\cdot 
  P_n(\eta\,;\bm{\lambda}+\bm{\delta})
  &=(n+2a)P_n(\eta\,;\bm{\lambda}),
\end{align}
and for the continuous dual Hahn case:
\begin{align}
  \phi_0(x\,;\bm{\lambda}+\bm{\delta})^{-1}\,X(\bm{\lambda})\,\,
  \phi_0(x\,;\bm{\lambda})\cdot P_n(\eta\,;\bm{\lambda})
  &=P_n(\eta\,;\bm{\lambda}+\bm{\delta}),\\
  \phi_0(x\,;\bm{\lambda})^{-1}\,X(\bm{\lambda})^{\dagger}\,\,
  \phi_0(x\,;\bm{\lambda}+\bm{\delta})\cdot 
  P_n(\eta\,;\bm{\lambda}+\bm{\delta})
  &=\!\!\prod_{1\leq j<k\leq 3}(n+a_j+a_k)\cdot P_n(\eta\,;\bm{\lambda}).
  \label{cdhxdagsim}
\end{align}
Therefore $X^{\dagger}$ ($X$) compensates the parameter shift caused by
$\mathcal{A}$ ($\mathcal{A}^\dagger$), so that the effect of $a^{(-)}$
($a^{(+)}$) is to give the polynomial with the same parameter
$\bm{\lambda}$ of degree one lower (higher). This result is new. 

Let us close this Appendix with a remark on the formal definition of the
annihilation-creation operators used within the 
framework of shape-invariant quantum mechanics \cite{coherents,os4}.
A unitary operator $\mathcal{U}$ ($\mathcal{U}^\dagger$) is defined
as a map between two orthonormal bases with neighbouring parameters,
$\{\hat{\phi}_n(x;\bm{\lambda})\}$ and 
$\{\hat{\phi}_n(x;\bm{\lambda}+\bm{\delta})\}$:
\begin{equation}
  \mathcal{U}\hat{\phi}_n(x\,;\bm{\lambda})\eqdef
  \hat{\phi}_{n}(x\,;\bm{\lambda}+\bm{\delta}),\quad
  \mathcal{U}^{\dagger}\hat{\phi}_{n}(x\,;\bm{\lambda}+\bm{\delta})
  =\hat{\phi}_n(x\,;\bm{\lambda}).
\end{equation}
This  allows to introduce new annihilation-creation operators
in a factorised form
\begin{equation}
  \hat{a}\eqdef\mathcal{U}^{\dagger}\mathcal{A}(\bm{\lambda}),\quad
  \hat{a}^{\dagger}=\mathcal{A}(\bm{\lambda})^{\dagger}\mathcal{U},
\end{equation}
which satisfy $\mathcal{H}=\hat{a}^{\dagger}\hat{a}/2=
\mathcal{A}(\bm{\lambda})^{\dagger}\mathcal{A}(\bm{\lambda})/2$.
The operator $\mathcal{U}$ is rather formal and it cannot be expressed 
as a differential or a difference operator. This operator $\mathcal{U}$
can be considered as unitarisation of the natural factorisation operator $X$
discussed above.

\section*{Appendix C: Some definitions related to the
hypergeometric and $q$-hypergeometric functions}
\renewcommand{\theequation}{C.\arabic{equation}}
\setcounter{equation}{0}

For reader's convenience we collect several definitions related to 
the ($q$-)hypergeometric functions\cite{koeswart}.

\noindent
$\circ$ Pochhammer symbol $(a)_n$ :
\begin{equation}
  (a)_n\eqdef\prod_{k=1}^n(a+k-1)=a(a+1)\cdots(a+n-1)
  =\Gamma(a+n)/\Gamma(a).
  \label{defPoch}
\end{equation}
$\circ$ $q$-Pochhammer symbol $(a\,;q)_n$ :
\begin{equation}
  (a\,;q)_n\eqdef\prod_{k=1}^n(1-aq^{k-1})=(1-a)(1-aq)\cdots(1-aq^{n-1}).
  \label{defqPoch}
\end{equation}
$\circ$ hypergeometric series ${}_rF_s$ :
\begin{equation}
  {}_rF_s\Bigl(\genfrac{}{}{0pt}{}{a_1,\,\cdots,a_r}{b_1,\,\cdots,b_s}
  \Bigm|z\Bigr)
  \eqdef\sum_{n=0}^{\infty}
  \frac{(a_1,\,\cdots,a_r)_n}{(b_1,\,\cdots,b_s)_n}\frac{z^n}{n!}\,,
  \label{defhypergeom}
\end{equation}
where $(a_1,\,\cdots,a_r)_n\eqdef\prod_{j=1}^r(a_j)_n
=(a_1)_n\cdots(a_r)_n$.\\
$\circ$ $q$-hypergeometric series (the basic hypergeometric series) 
${}_r\phi_s$ :
\begin{equation}
  {}_r\phi_s\Bigl(
  \genfrac{}{}{0pt}{}{a_1,\,\cdots,a_r}{b_1,\,\cdots,b_s}
  \Bigm|q\,;z\Bigr)
  \eqdef\sum_{n=0}^{\infty}
  \frac{(a_1,\,\cdots,a_r\,;q)_n}{(b_1,\,\cdots,b_s\,;q)_n}
  (-1)^{(1+s-r)n}q^{(1+s-r)n(n-1)/2}\frac{z^n}{(q\,;q)_n}\,,
  \label{defqhypergeom}
\end{equation}
where $(a_1,\,\cdots,a_r\,;q)_n\eqdef\prod_{j=1}^r(a_j\,;q)_n
=(a_1\,;q)_n\cdots(a_r\,;q)_n$.\\
$\circ$ Bessel function $J_a(z)$ :
\begin{equation}
  J_a(z)\eqdef\frac{(z/2)^a}{\Gamma(a+1)}\,
  {}_0F_1\Bigl(\genfrac{}{}{0pt}{}{-}{a+1}\Bigm|-\frac{z^2}{4}\Bigr).
  \label{defBes}
\end{equation}
$\circ$ Hermite polynomial $H_n(x)$ :
\begin{equation}
  H_n(x)\eqdef(2x)^n\,
  {}_2F_0\Bigl(\genfrac{}{}{0pt}{}{-n/2,\,-(n-1)/2}{-}\Bigm|
  -\frac{1}{x^2}\Bigr).
  \label{defHer}
\end{equation}
$\circ$ Laguerre polynomial $L^{(\alpha)}_n(x)$ :
\begin{equation}
  L^{(\alpha)}_n(x)\eqdef\frac{(\alpha+1)_n}{n!}\,
  {}_1F_1\Bigl(\genfrac{}{}{0pt}{}{-n}{\alpha+1}\Bigm|x\Bigr).
  \label{defLag}
\end{equation}
$\circ$ Jacobi polynomial $P^{(\alpha,\beta)}_n(x)$ :
\begin{equation}
  P^{(\alpha,\beta)}_n(x)\eqdef\frac{(\alpha+1)_n}{n!}\,
  {}_2F_1\Bigl(\genfrac{}{}{0pt}{}{-n,\,n+\alpha+\beta+1}{\alpha+1}
  \Bigm|\frac{1-x}{2}\Bigr),
  \label{defJac}
\end{equation}
which satisfies 
$P^{(\beta,\alpha)}_n(x)=(-1)^nP^{(\alpha,\beta)}_n(-x)$.\\
$\circ$ Gegenbauer polynomial $C^{(\lambda)}_n(x)$ :
\begin{equation}
  C^{(\lambda)}_n(x)\eqdef\frac{(2\lambda)_n}{(\lambda+1/2)_n}
  P^{(\lambda-1/2,\lambda-1/2)}_n(x).
  \label{defGege}
\end{equation}
$\circ$ Meixner-Pollaczek polynomial $P^{(a)}_n(x\,;\phi)$ :
\begin{equation}
  P^{(a)}_n(x\,;\phi)\eqdef\frac{(2a)_n}{n!}\,e^{in\phi}
  {}_2F_1\Bigl(\genfrac{}{}{0pt}{}{-n,\,a+ix}{2a}\Bigm|
  1-e^{-2i\phi}\Bigr).
  \label{defMP}
\end{equation}
$\circ$ continuous Hahn polynomial $p_n(x\,;a_1,a_2,a'_1,a'_2)$ :
\begin{align}
  p_n(x\,;a_1,a_2,a'_1,a'_2)&\eqdef
  i^n\frac{(a_1+a'_1)_n(a_1+a'_2)_n}{n!}\n
  &\qquad\times
  {}_3F_2\Bigl(\genfrac{}{}{0pt}{}{-n,\,n+a_1+a_2+a'_1+a'_2-1,\,a_1+ix}
  {a_1+a'_1,\,a_1+a'_2}\Bigm|1\Bigr), 
  \label{defcH}
\end{align}
which is symmetric under $a_1\leftrightarrow a_2$ and 
$a'_1\leftrightarrow a'_2$ separately.\\
$\circ$ continuous dual Hahn polynomial $S_n(x^2\,;a_1,a_2,a_3)$ :
\begin{equation}
  S_n(x^2\,;a_1,a_2,a_3)\eqdef
  (a_1+a_2)_n(a_1+a_3)_n\ 
  {}_3F_2\Bigl(\genfrac{}{}{0pt}{}{-n,\,a_1+ix,\,a_1-ix}
  {a_1+a_2,\,a_1+a_3}\Bigm|1\Bigr), 
  \label{defcdH}
\end{equation}
which is symmetric under the permutations of $(a_1,a_2,a_3)$.\\
$\circ$ Wilson polynomial $W_n(x^2\,;a_1,a_2,a_3,a_4)$ :
\begin{align}
  W_n(x^2\,;a_1,a_2,a_3,a_4)&\eqdef
  (a_1+a_2)_n(a_1+a_3)_n(a_1+a_4)_n\n
  &\qquad\times 
  {}_4F_3\Bigl(
  \genfrac{}{}{0pt}{}{-n,\,n+\sum_{j=1}^4a_j-1,\,a_1+ix,\,a_1-ix}
  {a_1+a_2,\,a_1+a_3,\,a_1+a_4}\Bigm|1\Bigr), 
  \label{defW}
\end{align}
which is symmetric under the permutations of $(a_1,a_2,a_3,a_4)$.\\
$\circ$ Askey-Wilson Hahn polynomial $p_n(\cos x\,;a_1,a_2,a_3,a_4|q)$ :
\begin{align}
  p_n(\cos x\,;a_1,a_2,a_3,a_4|q)&\eqdef
  a_1^{-n}(a_1a_2,a_1a_3,a_1a_4\,;q)_n\n
  &\qquad\times 
  {}_4\phi_3\Bigl(\genfrac{}{}{0pt}{}{q^{-n},\,a_1a_2a_3a_4q^{n-1},\,
  a_1e^{ix},\,a_1e^{-ix}}{a_1a_2,\,a_1a_3,\,a_1a_4}\Bigm|q\,;q\Bigr), 
  \label{defAW}
\end{align}
which is symmetric under the permutations of $(a_1,a_2,a_3,a_4)$.


\end{document}